\documentclass[11pt]{article}
\usepackage[letterpaper,margin=1in]{geometry}

\usepackage{amsmath,amssymb,amsthm,amsfonts,latexsym,bbm,xspace,graphicx,float,mathtools}
\usepackage{thm-restate}
\usepackage{bbm}
\usepackage{comment}
\usepackage[ruled, vlined, linesnumbered]{algorithm2e}

\usepackage{physics}
\usepackage{xcolor}
\usepackage{enumitem}

\usepackage{dsfont}

\usepackage{array}
\usepackage{tikz}
\tikzset{every picture/.style={line width=0.75pt}} 

\usepackage[pagebackref,colorlinks,citecolor=blue!35!black,linkcolor=magenta,bookmarks=true,unicode]{hyperref}
\hypersetup{
pdftitle={Clifford testing},
pdfauthor={Marcel Hinsche, Zongbo (Bob) Bao, Philippe van Dordrecht, Jens Eisert, Jop Bri\"{e}t, Jonas Helsen},
pdfkeywords={quantum information, quantum property testing, Clifford group}
}
 
\usepackage[nameinlink,capitalize]{cleveref}
\crefname{equation}{Eq.}{Eqs.}
\Crefname{equation}{Eq.}{Eqs.}

\newtheorem{thm}{Theorem}[section]
\newtheorem{lem}[thm]{Lemma}

\newtheorem{fact}[thm]{Fact}
\newtheorem{cor}[thm]{Corollary}

\theoremstyle{definition}
\newtheorem{definition}[thm]{Definition}
\newtheorem{remark}[thm]{Remark}
\newtheorem{example}[thm]{Example}

\newcommand{\C}{\mathbb{C}}
\newcommand{\ct}{^\dagger}
\newcommand{\tn}[1]{^{\otimes #1}}

\newcommand{\stab}{\mathrm{Stab}}

\newcommand{\Cln}{\mathrm{Cl}(n)}

\newcommand*{\poly}{\mathrm{poly}}
\newcommand{\F}{\mathbb{F}_2}
\newcommand{\Ft}{\mathbb{F}_2^t}
\newcommand{\Fn}{\mathbb{F}_2^n}

\newcommand{\E}{\mathbb{E}}

 \newcommand{\idx}{\mathbbm{1}}
 \newcommand{\Ftn}{\F^{2n}}

 \def\f{\frac}
 
 \def\cliff{\mathrm{Cliff}}
 
 \newcommand{\kett}[1]{\ket{ #1 \rangle\!}}
 \newcommand{\bbra}[1]{\bra{\!\langle #1}}
 
 \DeclarePairedDelimiter{\lrpt}{(}{)}
 \DeclarePairedDelimiter{\lrst}{\lbrace}{\rbrace}
 \DeclarePairedDelimiter{\lrbt}{[}{]}
 
 \def\p{\lrpt*}
 \def\lrs{\lrst*}
 \def\lrb{\lrbt*}
 
 \newcommand{\eps}{\varepsilon}
 \renewcommand{\epsilon}{\varepsilon}

\begin{document}

\title{Clifford testing: algorithms and lower bounds}
\date{\today}
 \author{Marcel Hinsche\thanks{Freie Universität Berlin, \texttt{m.hinsche@fu-berlin.de}}\and Zongbo (Bob) Bao\thanks{Centrum Wiskunde \& Informatica (CWI) and QuSoft, Amsterdam, \texttt{zongbo.bao@cwi.nl}}\and{Philippe van Dordrecht} \thanks{Centrum Wiskunde \& Informatica (CWI) and QuSoft, Amsterdam, \texttt{Philippe.Dordrecht@cwi.nl}}\and Jens Eisert \thanks{Freie Universität Berlin, \texttt{jense@zedat.fu-berlin.de}}
 \and Jop Bri{\"e}t \thanks{Centrum Wiskunde \& Informatica (CWI) and QuSoft, Amsterdam, \texttt{j.briet@cwi.nl}}  \and Jonas Helsen\thanks{Centrum Wiskunde \& Informatica (CWI) and QuSoft, Amsterdam, \texttt{jonas@cwi.nl}}}
\maketitle

\begin{abstract}
 We consider the problem of Clifford testing, which asks whether a black-box $n$-qubit unitary is a Clifford unitary or at least $\varepsilon$-far from every Clifford unitary. We give the first 4-query Clifford tester, which decides this problem with probability~$\poly(\varepsilon)$.  
 This contrasts with the minimum of 6 copies required for the closely-related task of stabilizer testing. 
 We show that our tester is tolerant, by adapting techniques from tolerant stabilizer testing to our setting. In doing so, we settle in the positive a conjecture of Bu, Gu and Jaffe, by proving a polynomial inverse theorem for a non-commutative Gowers 3-uniformity norm.
 We also consider the restricted setting of single-copy access, where we give an $O(n)$-query Clifford tester that requires no auxiliary memory qubits or adaptivity. We complement this with a lower bound, proving that any such, potentially adaptive, single-copy algorithm needs at least $\Omega(n^{1/4})$ queries. 
 To obtain our results, we leverage the structure of the commutant of the Clifford group, obtaining several technical statements that may be of independent interest. 
\end{abstract}

\section{Introduction}\label{sec:introduction}

Stabilizer testing---deciding whether an unknown state is close to a stabilizer state or far from every stabilizer state---has recently seen several remarkable advances~\cite{grossSchurWeylDuality2021, grewalImprovedStabilizerEstimation2024,iyerTolerantTestingStabilizer2024}.
While the task originates in quantum property testing \cite{montanaroSurveyQuantumProperty2016}, subsequent work has revealed deep connections to other areas of quantum information theory, mathematics and computer science. 
In particular, stabilizer testing is directly linked to the representation theory of the Clifford group \cite{grossSchurWeylDuality2021}, to the resource theory of magic \cite{buStabilizerTestingMagic2025, bittelOperationalInterpretationStabilizer2025} and quadratic Fourier analysis~\cite{ arunachalamPolynomialTimeTolerantTesting2025, baoTolerantTestingStabilizer2025,mehrabanImprovedBoundsTesting2025}.
These insights have led to steadily improving stabilizer testers, but also to surprising advances in classical algebraic property testing and algorithmic additive combinatorics~\cite{BrietCastroSilva2025QuadraticGoldreichLevin, ArunachalamCastroSilvaDuttGur2025AlgPFR}, as well as algorithms that operate in the restricted setting of single-copy access \cite{hinscheSingleCopyStabilizerTesting2025}.

In this work, we build on all of these advances and tackle the natural dynamic analog of stabilizer testing, namely \emph{Clifford testing}: given query access to an unknown $n$-qubit unitary $U$, determine whether it belongs to the Clifford group or is far from it.
Clifford testing has structural similarities to stabilizer testing, but as a form of unitary property testing it has some extra properties that make it theoretically  interesting in its own right. We will discuss some of these properties further down in the introduction, and point out informally how our work addresses these, with a more formal rundown of results given in \Cref{ssec:results}. We also discuss connections to the existing literature on stabilizer testing, as well as intriguing connections to additive combinatorics.

\paragraph{Clifford testing.} In the context of property testing, a natural way to measure proximity to the Clifford group is in terms 
of the \emph{Clifford fidelity}, 
\begin{equation}
F_{\mathrm{Cliff}}(U): = \max_{C\in\mathrm{Cl}(n)} 2^{-2n}\big|\mathrm{tr}(U\ct C)\big|^2,
\end{equation}
where $\mathrm{Cl}(n)$ denotes the $n$-qubit Clifford group.\footnote{For discussions of suitable distance measures in the context of property testing unitary operators and, more generally, quantum channels, see
Refs.~\cite{montanaroSurveyQuantumProperty2016,rosenthalQuantumChannelTesting2024}.}
This bears close resemblance 
to stabilizer fidelity (see~\cref{sec:stabfidelity}).
We say that~$U$ is \emph{$\eps$-far} from Clifford if $F_{\mathrm{Cliff}}(U)<1-\eps$ and \emph{$\eps$-close} otherwise. A quantum algorithm is a (one-sided) \emph{$\eps$-Clifford tester} if it accepts every Clifford unitary with probability at least~$2/3$ and rejects any unitary that is $\eps$-far from Clifford with probability at least~$2/3$.

\paragraph{Inverse-free Clifford testing.}
The first Clifford testers were considered by Low~\cite{lowLearningTestingAlgorithms2009} and Wang~\cite{wangPropertyTestingUnitary2011}. These testers however rely on access to the unitary $U$ \emph{and its inverse $U^{\dagger}$}, giving $\poly(n/\eps)$- and $O(1/\eps^2)$-query $\eps$-testers, respectively.
Access to $U^{\dagger}$ can be achieved in circuit-based models by reversing the circuit, assuming the gate set contains or can efficiently synthesize inverses. However, in many physical or experimental settings, $U$ represents the evolution of a device, or other process where implementing $U^{\dagger}$ would require reversing the system’s dynamics, which may be infeasible. This motivates the question of \emph{inverse-free Clifford testing}, which we will consider in this work. 

Gross, Nezami, and Walter~\cite{grossSchurWeylDuality2021} constructed a 6-query algorithm for stabilizer testing. They noted~\cite[Remark~3.7]{grossSchurWeylDuality2021}, that this can be adapted to Clifford testing via the Choi isomorphism. In this work, we make this connection precise by relating stabilizer fidelity to Clifford fidelity (see \Cref{sec:reduction-clifford-to-stabilizer-testing}), yielding an inverse-free 6-query Clifford tester.\\

Although stabilizer testing is known to require at least $6$ queries~\cite{damanikMScThesisAfstudeerscriptie,grossSchurWeylDuality2021}, it is not clear that  (inverse-free) Clifford testing should need $6$ queries.
Intuition for this is provided by the fact that the Clifford group fails to be a unitary $4$-design, meaning that it might in principle be possible to distinguish a Clifford from a non-Clifford unitary 
using only $4$ queries. Our first result confirms this by giving an inverse-free Clifford tester that uses only 4 (entangled) queries (\Cref{alg:4cltester}). 
This discrepancy with the stabilizer case is technically interesting, as the stabilizer states also fail to form an exact state $4$-design. However they do form an approximate additive error state $4$-design (in fact even  a $5$-design) with exponentially small additive error~\cite{grevinkWillItGlue2025}. Our result can be seen as showing that a similar approximate statement does not hold for the Clifford group. 

\paragraph{Tolerant Clifford testing.}

In this work we shall also be concerned with \emph{tolerant} testing, a natural extension of the one-sided paradigm of property testing~\cite{ParnasRonRubinfeld2006}.
In analogy with recent works on stabilizer testing~\cite{baoTolerantTestingStabilizer2025,arunachalamPolynomialTimeTolerantTesting2025,mehrabanImprovedBoundsTesting2025}, this is more naturally expressed in terms of fidelity.
For $1>\eps_1>\eps_2\geq 0$, a quantum algorithm is an \emph{$(\eps_1,\eps_2)$-tolerant Clifford tester} if, given an $n$-qubit unitary~$U$, it accepts with probability at least~2/3 if~$U$ is $F_{\mathrm{Cliff}}(U) \geq \eps_1$ and rejects with probability at least~2/3 if $U$ is $F_{\mathrm{Cliff}}(U) \leq \eps_2$. Our second result shows that our $4$-query Clifford tester is tolerant in this sense.

Our tolerant analysis, which extends the techniques used in tolerant stabilizer testing, has an interesting connection to the recent work of Bu, Gu \& Jaffe~\cite{buQuantumHighOrderAnalysis2025,buStabilizerTestingMagic2025}, which defines a non-commutative analogue of the famous Gowers uniformity norms from additive combinatorics. 
Such uniformity norms measure how much a function oscillates after it has been derived a number of times. 
Intuition from calculus suggests that small oscillations imply some sort of polynomial structure, and deep inverse theorems confirm this intuition (see for instance~\cite{GreenTao2008}).
Bu et al. conjecture in Ref.~ \cite{buQuantumHighOrderAnalysis2025} that such an inverse theorem holds for their non-commutative version of the $U^3$-norm. Our tolerant analysis resolves this conjecture in the positive by connecting the non-commutative $U^3$-norm directly to the acceptance probability of our $4$-query Clifford tester. This result fits in a recent trend of intriguing connections between these areas and quantum information theory~\cite{ AsadiGolovnevGurShinkarSubramanian2024QuantumWCAverageLinear, BrietEscuderoGutierrezGribling2024GrothendieckInequalities, BuGuJaffe2025QuantumRuzsaDivergence, BrietCastroSilva2025QuadraticGoldreichLevin, ArunachalamCastroSilvaDuttGur2025AlgPFR}.

\paragraph{Single-copy Clifford testing.}
Our other results pertain to resource-restricted query models for Clifford testing, which are motivated by the practical challenges of implementing testing algorithms.
In particular, we consider two key resource restrictions:
\begin{enumerate}[itemsep=1pt, topsep=5pt]
    \item Single-copy access (or \textit{incoherent access}, or operating \textit{without quantum memory}).
    \item Lack of an auxiliary system.
\end{enumerate}
The first restriction, single-copy access, has already received significant attention in quantum learning theory and property testing \cite{bubeckEntanglementNecessaryOptimal2020,aharonovQuantumAlgorithmicMeasurement2022,chenExponentialSeparationsLearning2022a,fawziQuantumChannelCertification2023, harrowApproximateOrthogonalityPermutation2023, chenOptimalTradeoffsEstimating2024,pauliMeasurements2025}. For state-related tasks, single-copy algorithms only process one copy of the state at a time, in contrast to multi-copy algorithms that can act jointly on several copies. For tasks involving unitaries or channels, single-copy algorithms are those that keep no quantum memory between queries: each round consists of preparing an input, applying the channel once, and measuring the entire output system.
The restriction to single-copy access is motivated by the technological difficulty of maintaining a coherent quantum memory or performing joint multi-copy operations. However, single-copy algorithms can exhibit dramatically increased sample complexities, often even exponentially, compared to the multi-copy setting. 

The second restriction, lack of an auxiliary system, arises more specifically in the context of learning and testing unitaries or channels. With access to an auxiliary register, an algorithm can prepare entangled inputs, send only part of the state through the channel, and then measure the entire joint system. This entanglement can provide a significant advantage. Following 
the nomenclature laid out in Ref.~\cite{rosenthalQuantumChannelTesting2024}, we refer to algorithms without such an auxiliary system as \emph{auxiliary-free} (ancilla-free), and to those that make use of it as \emph{auxiliary-assisted} (ancilla-assisted). 
\medskip

Here, we investigate Clifford testing in these resource-restricted query models. To construct single-copy Clifford testing algorithms, our starting point is the work  \cite{hinscheSingleCopyStabilizerTesting2025} which gives a single-copy stabilizer testing algorithm using $O(n/\epsilon^2)$ copies of the unknown state. In the auxiliary-assisted setting, by preparing copies of the Choi state and feeding them one at a time into this algorithm, one can obtain an auxiliary-assisted single-copy algorithm that inherits the complexity of the 
scheme in Ref.\ \cite{hinscheSingleCopyStabilizerTesting2025}. However, this tester is not auxiliary-free (as we need memory for the Choi states). We also give an auxiliary-free single-copy $\eps$-Clifford tester (which is substantially more difficult to derive) that uses $O(n/\epsilon^3)$ queries and time $O(n^3/\epsilon)$. Finally, we prove that any auxiliary-free tester requires~$\Omega(n^{1/4})$ queries, even when the tester is allowed to make \emph{adaptive} queries (which can made a qualitative difference in some scenarios \cite{rosenthalQuantumChannelTesting2024}).

\renewcommand{\arraystretch}{1.5}
\begin{table}[h!]
\small
\centering
\begin{tabular}{|>{\centering\arraybackslash}m{2.5cm}|
                >{\centering\arraybackslash}m{2.5cm}|
                >{\centering\arraybackslash}m{3.5cm}|
                >{\centering\arraybackslash}m{3.5cm}|}
    \hline
    & \textbf{Multi-copy} & \multicolumn{2}{c|}{\textbf{Single-copy}} \\ \hline
    &  & \textbf{Auxiliary-free} & \textbf{Auxiliary-assisted} \\ \hline
    \textbf{Clifford testing} & $ t =4 $ & $\Omega(n^{1/4}) \leq t \leq O(n)$ & Open \\ \hline
    \textbf{Stabilizer testing} & $t=6$ \cite{grossSchurWeylDuality2021} & \multicolumn{2}{c|}{$\Omega(n^{1/2}) \leq t \leq O(n)$ \cite{hinscheSingleCopyStabilizerTesting2025}}  \\ \hline
\end{tabular}
 \caption{Summary of our results: Upper and lower bounds on the query complexity for inverse-free Clifford testing and comparison to sample complexity of stabilizer testing.}\label{tab:overview-results}
\end{table}

\subsection{Summary of results} \label{ssec:results}

Below, we summarize our results in detail.
The first two results give performance guarantees for Clifford testing algorithms.

\begin{restatable}[One-sided 4-query Clifford tester]{thm}{fourtest}
\label{thm:fourtest}
    There exists a quantum algorithm that, given an $n$-qubit unitary~$U$, makes~4 queries to~$U$ and for any $\eps>0$, has the following completeness and soundness guarantees:
    \begin{itemize}
        \item It accepts if~$U$ is a Clifford unitary.
        \item It rejects with probability $ \min\big(\tfrac{1}{4}, \tfrac{\eps}{2}\big)$ if $F_{\mathrm{Cliff}}\left(U\right)\leq 1-\varepsilon$.
    \end{itemize}
\end{restatable}

\begin{restatable}[Two-sided 4-query Clifford tester]{thm}{tolerantfourtest}
\label{thm:tolerantfourtest}
There exists quantum algorithm that, given an $n$-qubit unitary~$U$, makes~4 queries to~$U$ and for any $\eps>0$, has the following completeness and soundness guarantees:
    \begin{itemize}
         \item It accepts with probability~$\poly(\eps)$ if $F_{\mathrm{Cliff}}(U) \ge \varepsilon$.
        \item It reject with probability~$1-\poly(\eps)$ if $F_{\mathrm{Cliff}}\left(U\right)\leq \varepsilon$.
    \end{itemize}
\end{restatable}

By repeating these testers $\poly(\eps)$ times, we obtain constant-query testers  with perfect (resp.\ constant) completeness and soundness (see \Cref{sec:4-query-tester}).
In proving the existence of a tolerant Clifford tester we also settle a conjecture due to \cite{buQuantumHighOrderAnalysis2025}, pertaining to a non-commutative generalization of the Gowers uniformity norms (see \cref{def:uniformity-norm}).

\begin{restatable}[Inverse theorem for the $Q^3$ norm]{thm}{Qinverse}
\label{thm:Qinverse}
    For any $n$-qubit unitary~$U$, we have that
    \begin{equation}
        F_{\mathrm{Cliff}}(U) \geq \poly\big(\|U\|_{Q^3}\big).
    \end{equation}
\end{restatable}

Finally, we prove upper and lower bounds for Clifford testing in the single-copy access model.

\begin{restatable}[Efficient auxiliary-free, single-copy Clifford tester]{thm}{auxfreesingle}
\label{thm:upper-bound-single-copy}
    There exists an auxiliary-free single-copy $\eps$-Clifford tester that uses $\tilde O(n/\epsilon^3)$ queries and time $\tilde O(n^3/\epsilon^2)$.
\end{restatable}

\begin{restatable}[Lower bound for auxiliary-free, single-copy Clifford testers]{thm}{lowerbound}
\label{thm:lower-bound}
     Any auxiliary-free single-copy algorithm for Clifford tester requires at least $\Omega(n^{1/4})$ queries.
\end{restatable}

This bound holds also against \textit{adaptive} algorithms which may choose input states and measurements for subsequent rounds based on measurement outcomes from previous round. Interestingly, we find that our proof technique for the lower bound does not straightforwardly extend to the auxiliary-assisted setting (see \cref{ssec:auxiliary-assisted-lower-bound} for more details).

\subsection{Technical overview}
\paragraph{From stabilizer testing to Clifford testing.}

To connect Clifford testing to stabilizer testing, we need to understand the relation between Clifford fidelity and stabilizer fidelity. Via the Choi–Jamiołkowski isomorphism, every unitary $U$ corresponds to its Choi state $\kett{U}$. 
We then observe that, since every Clifford Choi state is a stabilizer state,
\begin{equation}
    F_\mathrm{Cliff}(U) = \max_{C\in\mathrm{Cl}(n)}  |\langle\! \braket{C}{U}\!\rangle|^{2}\leq 
     \max_{S\in\stab(2n)}  |\braket{S}{U \rangle\!}|^2
     =
     F_{\stab}(\kett{U}).
\end{equation}
However, this one-sided inequality alone is insufficient to reduce Clifford testing to stabilizer testing.

Our first technical contribution resolves this by proving that the two fidelities are in fact polynomially equivalent (\cref{thm:relation-fstab-fcliff}),
\begin{equation}\label{eq:sandwich-cliff-stab}
 F_{\stab}(\kett{U})^6 \leq  F_{\cliff}(U) \leq F_{\stab}(\kett{U}),
\end{equation}
and that they even coincide whenever $ F_{\stab}(\kett{U})>1/2$. This sandwich inequality establishes a precise quantitative link between stabilizer and Clifford fidelity, thereby allowing Clifford testing to be reduced to stabilizer testing even in the tolerant sense. Importantly, the same inequality underlies the proof of \cref{thm:tolerantfourtest}, yielding performance guarantees for the novel 4-query tester.

\paragraph{Expected stabilizer fidelity and the auxiliary-free tester}
The idea behind the auxiliary-free single-copy algorithm is to sample a random $n$-qubit stabilizer state $\ket S$ and apply the
unknown unitary $U$ to prepare $U\ket{S}$. We then feed copies of $U\ket{S}$ into the single-copy stabilizer 
tester from Ref.\ \cite{hinscheSingleCopyStabilizerTesting2025}. Intuitively, since $\ket{S}$ is drawn at random, we should have a good chance that any non-Cliffordness in $U$ translates to non-stabilizerness of $U\ket{S}$. Our technical contribution here is to show that this strategy indeed works: 
We demonstrate that if~$U$ has Clifford fidelity
$1-\varepsilon$, the resulting state
$U\ket S$ will with probability $\Omega\left(\varepsilon\right)$ have
stabilizer fidelity $1-\Omega\left(\varepsilon\right)$
and can hence be tested by the single-copy stabilizer testing algorithm. 
To this end, we prove a strong sandwich inequality between Clifford and expected stabilizer fidelity (\cref{thm:average-stabilizer-fidelity}):
\begin{equation}
F_{\mathrm{Cliff}}(U)\leq \underset{\ket S\in\mathrm{Stab}\left(n\right)}{\mathbb{E}}\big [F_{\mathrm{Stab}}\left(U\ket S\right)\big] \leq \bigg[\frac{1}{8}F_{\mathrm{Stab}}(\kett{U}) + \frac{7}{8}+ O(2^{-n})\bigg]^{1/4}
\end{equation}
where, again, $F_{\mathrm{Stab}}(\kett{U}) = F_{\mathrm{Cliff}}(U)$ whenever  $F_{\mathrm{Cliff}}(U)>1/2$). We believe this sandwich inequality is of independent interest.

\paragraph{Clifford group forms an approximate unitary design for PPT operators}

To prove our lower bound on auxiliary-free single-copy Clifford testers in \cref{thm:lower-bound}, we analyze the ability of such testers to distinguish the $t$-fold Haar and Clifford twirls. Our key technical contribution here is a new structural statement about the Clifford group.

\begin{thm}[Clifford group is an approximate $t=o(n^{1/4})$-design for PPT operators]\label{thm:Clifford-design}
Let $\Phi_H^{(t)}= \mathbb{E}_{U\sim \mu_H} [U^{\otimes t}(\cdot)U^{\dagger,\otimes t}]$ be the $t$-fold Haar twirling channel and $\Phi_C^{(t)}=  \mathbb{E}_{C\sim \Cln}
[C^{\otimes t}(\cdot)C^{\dagger,\otimes t}]$ be the $t$-fold Clifford twirling channel. Then,
\begin{equation}
    \max_{\rho, M \in \mathrm{PPT}} \left| \tr\big (M\, \Phi_H^{(t)}(\rho)\big) - \tr\big(M \,\Phi_C^{(t)}(\rho)\big)  \right| \leq 2^{-n +O(t^4)}.
\end{equation}
\end{thm}
Here, $\mathrm{PPT}$ (positive partial transpose) denotes the set of operators that remain positive semidefinite under all partial transpositions across the $t$ copies; in particular, this set includes all product and separable operators.

This result can be viewed as showing that the Clifford group forms an approximate unitary design when the distinguishability metric is restricted to PPT operators---a relaxation of the usual diamond norm.
Apart from our main application in the single-copy lower bound, this result also finds an application in the recent work \cite{kretschmerDemonstratingUnconditionalSeparation2025}, which establishes an unconditional quantum–classical separation in memory usage. In their argument, a key step involves showing that an anti-concentration-type quantity, $\mathbb{E}_{C\sim\Cln}|\bra{\psi} C \ket{0^n}|^{2t}$, is well-approximated by its Haar-averaged counterpart.

To show \cref{thm:Clifford-design}, we analyze the Clifford commutant,
\begin{equation}
\mathrm{Comm}(\Cln, t) := \{A \in \mathcal{L}((\mathbb{C}^{2})^{\otimes n})^{\otimes t}) \;|\; [A, C^{\otimes t}] = 0 \ \forall C \in \Cln \},
\end{equation}
i.e., the space of operators that commute with all $C^{\otimes t}$ for $C \in \Cln$, which is precisely the subspace onto which the $t$-fold Clifford twirl projects.
Since we restrict the distinguishability metric to PPT operators, it is essential to understand the behavior of commutant generators under partial transposition.
In previous work \cite{hinscheSingleCopyStabilizerTesting2025}, it was shown that every nontrivial generator $R(T)$ of the Clifford commutant admits a non-unitary partial transpose. 
Here, we continue this line of study and complement it by showing that each generator also admits a unitary partial transpose\footnote{A similar unitarity result was recently obtained in Ref.\  \cite{bittelOperationalInterpretationStabilizer2025} using a completely different approach.} (\cref{thm:unitary-partial-transposes-rT}), which can be found efficiently in the number of copies $t$.

Our proof leverages the characterization of the Clifford commutant generators in terms of self-dual binary codes from \cite{grossSchurWeylDuality2021} and establishes a connection to matroid theory: by viewing the generator matrices of these codes as matroids, we can apply matroid intersection results, most notably Rado's theorem \cite{oxleyMatroidTheory2011}, to establish the existence of the desired partial transpose.

\subsection{Organization of this work}
The rest of this paper is organized as follows. In \cref{sec:preliminaries}, we collect background material from stabilizer testing theory and review the characterization of the Clifford commutant from \cite{grossSchurWeylDuality2021}.
In \cref{sec:reduction-clifford-to-stabilizer-testing}, we discuss the reduction from Clifford testing to stabilizer testing and prove the sandwich inequality \cref{eq:sandwich-cliff-stab}.
In \cref{sec:4-query-tester}, we present and analyze the 4-query Clifford tester and prove \cref{thm:fourtest}, \cref{thm:tolerantfourtest}, and \cref{thm:Qinverse}.
In \cref{sec:single-copy-upper-bound}, we demonstrate our auxiliary-free single-copy Clifford tester and formally prove \cref{thm:upper-bound-single-copy}.
Lastly, in \cref{sec:single-copy-lower-bounds}, we formally prove the single-copy lower bound from \cref{thm:lower-bound} by proving \cref{thm:Clifford-design}.

\subsection*{Author Contributions}
MH and JH conceived of the project and derived the main results. MH wrote the bulk of the manuscript, with input from JH. ZB developed and wrote down the algorithm in \cref{sec:algorithmic-proof}, which led up to the current proof of \Cref{thm:unitary-partial-transposes-rT}.
JB, PvD, and ZB contributed \cref{thm:tolerantfourtest} and \cref{thm:Qinverse}. All authors participated in finalizing the manuscript. 
\subsection*{Acknowledgements}
JH acknowledges funding from the Dutch Research Council (NWO) through a Veni grant (grant No.VI.Veni.222.331) and the Quantum Software Consortium (NWO Zwaartekracht Grant No.024.003.037). ZB is supported by the Dutch Ministry of Economic Affairs and Climate Policy (EZK), as part of the Quantum Delta NL programme (KAT2).
PvD is supported by the Dutch Research Council (NWO), as part of the NETWORKS programme (Grant No.~024.002.003) and the Dutch Ministry of Economic Affairs and Climate Policy (EZK), as part of the Quantum Delta NL programme.
JB is supported by the Dutch Research Council (NWO), as part of the NETWORKS programme (Grant No.~024.002.003).
The Berlin team has been supported by the BMFTR (QuSol, Hybrid++, DAQC), the DFG (SPP 2514, CRC 183), the Clusters of Excellence MATH+ and ML4Q, 
the Quantum Flagship (Millenion, PasQuans2),
and the European Research Council (Debuqc).

\section{Preliminaries}\label{sec:preliminaries}
In this section we set notation and recall a variety  of known facts about stabilizer states, the Clifford group, and stabilizer testing. This section is meant mostly for later reference, and can be skipped by readers familiar with the relevant material.\\

We begin by setting some notation. For a positive integer $n$, we define $\left[n\right]:=\left\{ 1,\dots,n\right\} $. We denote by $\F$
the finite field of $2$ elements and by $\Fn$ the
$n$-dimensional vector space over this field.

For $p$ a distribution over $\Fn$
and $V\subseteq\Fn$ a subset, we define the
\emph{weight} of~$V$ under~$p$ by $p\left(V\right):=\sum_{x\in V}p\left( x\right)$.
For any unitary $U$ on $n$ qubits, we use $\kett{U}$ to denote its Choi state, $\kett{U} = \left(U\otimes I\right)\ket{\Omega}$ where $\ket{\Omega}=\frac{1}{2^{n/2}} \sum_{x\in \F^n} \ket{x,x}$ denotes the maximally entangled state on $2n$ qubits.

\subsection{Inner products over \texorpdfstring{$\mathbb{F}_{2}$}{F2}}
In this work, we deal with binary vector spaces and two different inner products on them. These will feature in different contexts: The first is the standard inner product, which will feature in our discussion of the commutant of the $t$-fold tensor power action of the Clifford group.

\begin{definition}[Standard inner product]
 For $x, y\in\mathbb{F}_{2}^{t}$, we define their \emph{standard
inner product} as 
\begin{equation}
x\cdot y = x_{1}y_{1}+\dots+x_{t}y_{t}
\end{equation}
where operations are performed over $\mathbb{F}_{2}$. 
\end{definition}

\begin{definition}[Dual of subspace]
 Let $D\subseteq\mathbb{F}_{2}^{t}$ be a subspace. The \emph{dual} of $D$, denoted $D^{\perp}$, is defined as
\begin{equation}
D^{\perp}:=\left\{x\in\mathbb{F}_{2}^{t}:x\cdot y=0\,, \; \forall\,y \in D\,\right\} .
\end{equation}
\end{definition}

\begin{definition}[Self-orthogonal subspace, self-dual subspace]
    A subspace $D\subseteq\mathbb{F}_{2}^{t}$ is \textit{self-orthogonal}, if $D\subseteq D^\perp$. Furthermore, $D$ is \textit{self-dual} if $D=D^\perp$.
\end{definition}

In the context of the stabilizer formalism and its phase-space description in terms of Weyl operators, we will instead use the symplectic inner product.

\begin{definition}The \textit{symplectic inner product} between two vectors $x,y\in \F^{2n}$ is the bilinear form
\begin{equation}
  [x,y]=a\cdot b'+ a'\cdot b,
\end{equation}
where $x=(a,b), y=(a',b')$ and $a,b,a',b'\in \F^n$.
\end{definition}

\begin{definition}[Isotropic and Lagrangian subspace]
    A set $V\subseteq \F^{2n}$ is \textit{isotropic} if for all $x,y\in V$, we have that $[x,y]=0$. 
    If~$V$ is a subspace, then it \emph{Lagrangian} if it has dimension~$n$ (which is maximal).
\end{definition}

\subsection{Weyl operators and stabilizer states}
\label{sec:weyl}

We recall some well-known facts about stabilizer formalism.
The single-qubit Pauli matrices are denoted by $\left\{ I,X,Y,Z\right\} $.
The $n$-qubit Pauli group $\mathcal{P}_{n}$ is the set $\left\{ \pm1,\pm i\right\} \cdot\left\{ I,X,Y,Z\right\} ^{\otimes n}$.
The Clifford group is the normalizer of the Pauli group. We denote
the $n$-qubit Clifford group by $\Cln$. 

A pure $n$-qubit state is a \emph{stabilizer state} if there exists an Abelian
subgroup $S\subset\mathcal{P}_{n}$ consisting of $2^{n}$ Pauli operators $P\in\mathcal{P}_{n}$
(with phase of $+1$ or $-1$) such that
\begin{equation}
S=\left\{ P\in\mathcal{P}_{n}:P\ket{\psi}=\ket{\psi}\right\}.
\end{equation}
This Abelian group is the \emph{stabilizer group} of the stabilizer
state and determines it uniquely.
We denote stabilizer
states by $\ket S$ and denote the set of all pure $n$-qubit stabilizer states by $\mathrm{Stab}\left(n\right)$. 

An important subset of $2n$-qubit stabilizer states is formed by Choi states of Clifford unitaries.

\begin{lem}\label{lem:ClStab}
    For any $n$-qubit Clifford unitary $C\in \Cln$, we have that $\kett{C}\in \stab(2n)$.
\end{lem}

\begin{proof}
    The maximally entangled state~$\ket{\Omega}$ is a stabilizer state.
    Let $S\subseteq \mathcal P_{2n}$ be its stablizer group.
    Since $C\otimes I$ is a $2n$-qubit Clifford unitary, it follows that $(C\otimes I)S(C^\dagger\otimes I)$ is an Abelian group of size~$|S|=2^{2n}$ that stabilizes $\kett{C} = (C\otimes I)\ket{\Omega}$.
\end{proof}

We will refer to the Hermitian (unsigned) $n$-qubit Pauli operators in $\left\{ I,X,Y,Z\right\}^{\otimes n}$ as Weyl operators and label them via bitstrings of length $2n$ as follows:
\begin{definition}[Weyl operator]
For $x=(a,b)\in \F^n\times \F^n =\F^{2n} $, the Weyl operator $P_x$ is defined as
\begin{equation}
P_x=i^{a\cdot b} X^{a}Z^{b} = i^{a\cdot b} (X^{a_1} Z^{b_1})\otimes \cdots \otimes (X^{a_n} Z^{b_n} ) .
\end{equation}
\end{definition}
\noindent Here, as an exception, the inner product $a\cdot b$ on the phase in front is understood as being an
integer resulting from the inner product of two binary integer-vectors.

The Weyl operators $P_x$ form an orthogonal operator basis with respect to the trace inner product.
Define the ``Fourier coefficients'' of an $n$-qubit operator~$A$ by $\widehat{A}(x) = \tr(A P_x)/2^n$.
Then, we have the usual Fourier inversion formula
\begin{equation}\label{eq:fourier_inversion}
    A = \sum_{x\in \Ftn}\widehat{A}(x) P_x 
\end{equation}
as well as Parseval's identity
\begin{equation}
    \tr(A B^{\dagger}) = \sum_{x\in \F^{2n}}\widehat{A}(x)\overline{\widehat{B}(x)}.
\end{equation}
It follows that the Frobenius norm (or Hilbert-Schmidt norm) of~$A$ satisfies
\begin{equation}\label{eq:parseval}
    \lVert A\rVert_{2}^2 = 
    2^n\sum_{x \in \Ftn} |\widehat{A}(x)|^2.
\end{equation}

We will occasionally identify binary vector spaces and sets of Weyl operators.
By considering the unsigned Weyl operators corresponding to the Pauli operators forming a stabilizer group, every stabilizer
group can be uniquely associated to a Lagrangian subspace $M\subset \Ftn$. That is, Lagrangian
subspaces are in a one-to-one correspondence with unsigned stabilizer
groups.

\subsection{The characteristic distribution}
Next, we introduce the characteristic distribution associated to an $n$-qubit state.
\begin{definition}[Characteristic distribution of a state, \cite{grossSchurWeylDuality2021, grewalImprovedStabilizerEstimation2024}]
Let $\ket{\psi}$ be an $n$-qubit pure state. Then its corresponding characteristic distribution $p_{\ket{\psi}}$ is defined via
\begin{equation}
    p_{\ket{\psi}}(x)=2^{-n}|\langle \psi|P_x|\psi\rangle|^2.
\end{equation}
\end{definition}

Next, we gather some properties of the characteristic distribution.
\begin{fact}
    \label{fact:property-ppsi}
    Let $\ket{\psi}$ be an $n$-qubit pure state. Then, the characteristic distribution satisfies:
    \begin{enumerate}
        \item $\sum_{x\in \F^{2n}}p_{\ket{\psi}}(x)=1$.
        \item For all $x\in \F^{2n}$, $p_{\ket{\psi}}(x)\leq 2^{-n}$.
    \end{enumerate}
\end{fact}

\begin{fact}[Uncertainty principle, Lemma 3.10 in Ref.\ \cite{grossSchurWeylDuality2021}]
\label{fact:uncertainty-principle}Let $\ket{\psi}$ be an $n$-qubit
pure state. 
Then, the set
$\left\{ x\in\mathbb{F}_{2}^{2n}:2^n\:p_{\ket{\psi}}\left(x\right)>\frac{1}{2}\right\}$ is isotropic.
\end{fact}

The following lemma was  originally proved over the real numbers in Ref.\ \cite{Samorosnitsky2007}.
A straightforward proof for this appeared as the proof of \cite[Lemma 5.9]{iyerTolerantTestingStabilizer2024}.
\begin{lem}[Affine subspaces carry no more weight than their underlying subspace]\label{lem:removeshifts}
    Let $\ket{\psi}$ be an $n$-qubit state and $V\subseteq \F^{4n}$ be a subspace. Then, for any affine shift $(x,y)\in \F^{4n}\backslash V$, it holds that
    \begin{align}
        p_{\ket{\psi}}(V)\geq p_{\ket{\psi}}(V+(x,y)).
    \end{align}
    where $V+(x,y)=\{(v+x,w+y):(v,w)\in V\}$.
\end{lem}

\subsection{Stabilizer fidelity and Clifford fidelity}
\label{sec:stabfidelity}

\begin{definition}[Stabilizer fidelity]\label{def:stabilizer-fidelity}
    Let $\ket{\psi}$ be a pure $n$-qubit quantum state. Then, the \textit{stabilizer fidelity} of $\ket{\psi}$ is defined as:
\begin{equation}
    F_{\stab}(\ket{\psi}) 
    = \max_{\ket{S}\in \stab(n)} \abs{\braket{S}{\psi}}^2.
\end{equation}
\end{definition}

And we also recall the Clifford fidelity here.

\begin{definition}[Clifford fidelity]
Let $U\in \mathrm{U}\left(2^n\right)$ be an $n$-qubit unitary operator. The \emph{Clifford
fidelity} of $U$ is defined as 
\begin{equation}\label{eq:clifford-fidelity}
F_{\mathrm{Cliff}}(U)
=\max_{C\in\mathrm{Cl}(n)} 2^{-2n} \big|\mathrm{Tr}(U\ct C)\big|^2 .
\end{equation}
\end{definition}

The characteristic distribution of a state is closely related to its stabilizer fidelity. In particular, the following lemma states that the weight of the characteristic distribution on any isotropic subspace is a lower bound for the stabilizer fidelity.

\begin{fact}[Lower bound for stabilizer fidelity, proof of Theorem 3.3 of \cite{grossSchurWeylDuality2021}]\label{fact:stab-fidelity-lower-bound}
Let $\ket{\psi}$ be an $n$-qubit pure state and let $M\subset \Ftn$ be a Lagrangian subspace. Then, 
    \begin{equation}
      F_{\stab}\p{\ket{\psi}}\geq p_{\ket{\psi}}(M).
    \end{equation}
\end{fact}

We can also give upper bounds on the stabilizer fidelity and the Clifford fidelity in terms of the characteristic distribution.

\begin{lem}[Upper bound on the stabilizer fidelity, Lemma 4.2 of
Ref.\ \cite{grewalImprovedStabilizerEstimation2024}]    \label{lem:upperbound-stabilizer-fidelity}
    Let $\ket{\psi}$ be an $n$-qubit quantum state. Then,
    \begin{equation}
        \max p_{\ket{\psi}}(M) \ge F_{\stab}(\ket{\psi})^2,
    \end{equation}
    where the maximum is over all Lagrangian subspaces $M\subset \Ftn$.
\end{lem}

\begin{cor}[Upper bound for the stabilizer fidelity and the Clifford fidelity]
    \label{cor:upperbound-fidelity}
    For any $n$-qubit quantum state $\ket{\psi}$, we have
    \begin{equation}
       2^n \lVert p_{\ket{\psi}} \Vert_2^2 
       \ge F_{\stab}(\ket{\psi})^4.
    \end{equation}
\end{cor}

\begin{proof}
    For any $n$-qubit quantum state $\ket{\psi}$ and let $M\subset \F^{2n}$ be the Lagrangian attaining the maximum weight in \cref{lem:upperbound-stabilizer-fidelity}. Then,
    \begin{align}
        2^n\sum_{x\in \F^{2n}} p_{\ket{\psi}}(x)^2 &= 
          \p{2^n\sum_{x\in \Ftn} p_{\ket{\psi}}(x)^2} \p{\f{1}{2^n}\sum_{x\in \Ftn}\idx_M(x)^2} \\
          \nonumber
        &=  \p{\sum_{x\in \Ftn} p_{\ket{\psi}}(x)^2} \p{\sum_{x\in \Ftn}\idx_M(x)^2} \\
        \nonumber
        &\ge \p{\sum_{x\in\Ftn} p_{\ket{\psi}}(x) \idx_M(x)}^2 \\
        \nonumber
        &= p_{\ket{\psi}}(M)^2 \ge F_{\stab}(\ket{\psi})^4,
        \nonumber
    \end{align}
    where the first inequality uses Cauchy-Schwarz and the second follows from
        \Cref{lem:upperbound-stabilizer-fidelity}.
\end{proof}

\subsection{Commutant of Clifford tensor powers}\label{ssec:clifford-commutant}
In this section we recall a number of standard facts about the Clifford group and its $t$-fold tensor product representation (in particular the generators of the associated commutant). We first discuss the commutant for arbitrary $t$, recalling several known properties from the literature that will be used in deriving our single-copy lower bound in \cref{sec:single-copy-lower-bounds}. Then, we provide a slightly more detailed exposition of the $t=4$ commutant that will feature both in the analysis of our 4-query Clifford tester in \cref{sec:4-query-tester} as well as our auxiliary-free single-copy Clifford tester in \cref{sec:single-copy-upper-bound}.

We consider the commutant of the $t$-fold tensor power action of the Clifford group $\Cln$. That is, we study the space of linear operators on $((\mathbb{C}^{2})^{\otimes n})^{\otimes t}$---corresponding to $t$ copies of an $n$-qubit system---that commute with $C^{\otimes t}$ for all $C \in \Cln$. Formally, we define it as follows:
\begin{definition}[Commutant of $t$-th Clifford tensor power action]
    We define $\mathrm{Comm}(\Cln, t)$ as follows
    \begin{equation}
        \mathrm{Comm}(\Cln, t) := \{A \in \mathcal{L}(((\mathbb{C}^{2})^{\otimes n})^{\otimes t}) \;|\; [A, C^{\otimes t}] = 0 \quad \forall C \in \Cln \}.
    \end{equation}
\end{definition}
The seminal work \cite{grossSchurWeylDuality2021} characterized this commutant in terms of so-called stochastic Lagrangian subspaces:

\begin{definition}[Stochastic Lagrangian subspaces]\label{def:sigma_tt}
The set $\Sigma_{t,t}$ denotes the set of all subspaces $T\subset \mathbb{F}^{2t}_2$ with the following properties:
\begin{enumerate}
    \item Total isotropy:  $ x \cdot  x =  y \cdot  y \mod 4 $ for all $( x,  y) \in T$,
    \item Maximality: $\dim (T) = t$,
    \item $ 1_{2t} = (1,\dots,1) \in T$.
\end{enumerate}
We refer to elements in $\Sigma_{t,t}$ as \textit{stochastic Lagrangian subspaces}.
    
\end{definition}

The key result of 
Ref.\ \cite{grossSchurWeylDuality2021} is that the commutant $\mathrm{Comm}(\Cln, t)$ is spanned by operators associated with the stochastic Lagrangian subspaces $T\in \Sigma_{t,t}$.

\begin{thm}[Theorem 4.3 in Ref.\ \cite{grossSchurWeylDuality2021}]\label{thm:commutant-basis-gnw}
If $n\geq t-1$, then $\mathrm{Comm}(\Cln, t)$ has a basis given by the operators $R(T) := r(T)^{\otimes n}$, where $T\in\Sigma_{t,t}$ and 
\begin{equation}
    r(T) := \sum_{( x,  y) \in T} \ket{x}\!\bra{y}.
\end{equation}
\end{thm}

We note that recently Ref. \cite{bittelCompleteTheoryClifford2025} provided a different and complementary perspective on the basis $\{R(T)\}_{T\in \Sigma_{t,t}} $ in terms of so-called \textit{Pauli monomials}. In this work, we stick to the description in terms of stochastic Lagrangian subspaces.
Next, we collect several additional facts about the commutant here, with proofs found in other works:
\begin{fact}[Cardinality of $\Sigma_{t,t}$, Theorem 4.10 in Ref.\ \cite{grossSchurWeylDuality2021}]\label{fact:cardinality-sigma-tt}
\begin{equation}
        \left|\Sigma_{t,t}\right|=\prod_{k=0}^{t-2}\left(2^{k}+1\right)\leq2^{\frac{1}{2}(t^{2}+5t)} \,.
\end{equation}
\end{fact}

Similar to the approach taken in Ref.~\cite{harrowApproximateOrthogonalityPermutation2023}, we want to quantify the orthogonality of the operators $R(T)$ spanning the commutant $\mathrm{Comm}(\Cln, t)$. To this end, we define their corresponding Gram matrix as follows:

\begin{definition}[Gram matrix $G$ corresponding to $\Sigma_{t,t}$]
We define the Gram matrix corresponding to $\{R(T)\}_{T\in\Sigma_{t,t}}$ as the $|\Sigma_{t,t}| \times |\Sigma_{t,t}|$-matrix with entries given by
\begin{equation}
        G^{(n,t)}_{T,T'} := \trace \left( R(T)^{\dagger} R(T') \right) \quad \text{for }T,T'\in\Sigma_{t,t} \,.
\end{equation}
\end{definition}
We also define the Weingarten matrix $W^{(n,t)}$ with entries $W^{(n,t)}_{T,T'}$ as the (Moore-Penrose pseudo-) inverse of $G^{(n,t)}$. With this, we can expand the projector onto the $t$-th order Clifford commutant in terms of the generators as follows,

\begin{equation}\label{eq:commutant_expansion}
     \underset{C\sim \Cln}{\mathbb{E}}[C^{\otimes t}(\cdot)C^{\dagger\,\otimes t}] = \sum_{T,T'\in \Sigma_{t,t}} W^{(n,t)}_{T, T'}\;   
        \tr \big[ R(T')^{\dagger} (\cdot) \big]\; R(T)  .
\end{equation}

For convenience, we will usually drop the superscript $(n,t)$ on $W^{(n,t)}$. We will need a few facts about the entries of the Weingarten matrix in the limit of many qubits (holding $t$ fixed):
\begin{fact}[Weingarten asymptotics, \cite{haferkampEfficientUnitaryDesigns2023,helsenThriftyShadowEstimation2023}]\label{fact:Weingarten asymptotics}
For all $T \in \Sigma_{t,t}$ we have
\begin{align}
\left|W^{(n,t)}_{T,T}-2^{-nt}\right| \leq2^{-n(t+1)+O(t^{2})} ,
\end{align}
and for all $T\neq T' \in \Sigma_{t,t}$ we have
\begin{align}
\left|W^{(n,t)}_{T,T'}\right| \leq2^{-n(t+1)+O(t^{2})}.
\end{align}
\end{fact}

Ref.~\cite{grossSchurWeylDuality2021} further characterized the commutant by uncovering an important group structure within $\Sigma_{t,t}$ that captures the unitary sector of the generators $\{R(T)\}_{T\in \Sigma_{t,t}}$. To describe this, we introduce the following definition.
\begin{definition}[Stochastic orthogonal group $\mathrm{O}_t^{(1)}$]
The stochastic orthogonal group, denoted $\mathrm{O}_t^{(1)}$, is defined as the group of $t\times t$ binary matrices $O$ such that 
\begin{equation}\label{eq:preserving-mod4}
    O  x \cdot O x =  x \cdot  x \mod 4, \qquad \forall  x \in \Ft \,.
\end{equation}
\end{definition}

For any $O\in \mathrm{O}_t^{(1)}$, the subspace $T_O = \{(O x,  x) \, | \,  x \in \Ft \}$ is a stochastic Lagrangian subspace.
That is, $T_O \in \Sigma_{t,t}$ for all $O \in \mathrm{O}_t^{(1)}$. In the following, we will thus view $\mathrm{O}_t^{(1)}$ as a subset of $ \Sigma_{t,t}$, i.e., $\mathrm{O}_t^{(1)} \subset  \Sigma_{t,t}$.
Furthermore, we will denote the identity element in $\mathrm{O}_t^{(1)}$ and its subgroups by $e$, it corresponds to the diagonal subspace $\{ ( x,  x) \,| \,  x \in \mathbb{F}_2^t \}\in \Sigma_{t,t}$.
Notice also that the symmetric group on $t$ elements, denoted $\mathcal{S}_t$, can be viewed as a subgroup of $\mathrm{O}_t^{(1)}$ by considering its matrix representation on $\mathbb{F}_2^t$. Hence, we have the following chain of inclusions:
\begin{equation}\label{eq:inclusions}
    \mathcal{S}_t  \subset \mathrm{O}_t^{(1)}  \subset  \Sigma_{t,t}. 
\end{equation}
\noindent Some remarks on these inclusions:
\begin{itemize}
    \item For $t=3$, all three sets coincide.
    \item For $t=4,5$, $\mathcal{S}_t = \mathrm{O}_t^{(1)} $ while $\mathrm{O}_t^{(1)}$ is strictly contained in $\Sigma_{t,t}$.
    \item For $t\ge 6$, all three sets differ and both inclusions are strict.
\end{itemize}

\paragraph{The \texorpdfstring{$t=4$}{t=4} commutant.}

Finally, we recall some specifics about the
$t=4$ commutant of the Clifford group from \cite{zhuCliffordGroupFails2016,grossSchurWeylDuality2021}. Namely, $\Sigma_{4,4}$ is strictly larger than $\mathcal{S}_4 = \mathrm{O}_4^{(1)}$. The additional generators in $\Sigma_{4,4}$ can be written in terms of the following projector:
\begin{equation}\label{eq:pi_4_definition}
\Pi_{4}=\frac{1}{2^{2n}} \sum_{x \in \mathbb{F}_2^{2n}}P_{x}^{\otimes4}.
\end{equation}
This is a projector onto a subspace $V_{n,4}\subset ((\mathbb{C}^2)^{\otimes n})^{\otimes 4}$ of dimension $2^{(t-2)n}=2^{2n}$ which is also a CSS stabilizer code. The projector is proportional to a specific generator of the commutant~\cite{grossSchurWeylDuality2021}
\begin{equation}
R\left(T_{4}\right)=2^n\Pi_{4}.
\end{equation}

It follows that for all Clifford unitaries $C\in\mathrm{Cl}\left(n\right)$
\begin{equation}
\left[\Pi_{4},C^{\otimes4}\right]=0.
\end{equation}

\noindent An orthonormal basis for this CSS code space is given by tensor products
of Bell states 
\begin{equation}\label{eq:basis}
V_{n,4}=\mathrm{span}\{ \ket{P_{x}\rangle\!}^{\otimes2}=\left(P_{x}\otimes I\ket{\Omega}\right)^{\otimes2}\:|\:x\in\mathbb{F}_{2}^{2n}\}.
\end{equation}
Hence, we can write $\Pi_4$ as
\begin{equation}
    \Pi_4 = \sum_{x\in\Ftn} \kett{P_x}\bbra{P_x} \otimes  \kett{P_x}\bbra{P_x}.
\end{equation}
It follows that the 2-outcome POVM $\left\{ \Pi_{4},I-\Pi_{4}\right\} $
can be realized by measuring in the Bell basis. Finally, we can express $\mathbb{E}_{\ket S \sim \stab(4)}\ket S\bra S^{\otimes4}$ via $\Pi_{4}$
as follows:
\begin{fact}[c.f.\ Corollary 1 in 
Ref.\ \cite{zhuCliffordGroupFails2016}]
 \label{fact:four-copy-average-stabilizer-state}Let $\Pi_{\mathrm{sym}}$
be the projector 
onto the symmetric subspace of $\left(\mathbb{C}^{d}\right)^{\otimes4}$,
i.e.,  $\mathrm{Sym}_{4}$$\left(\mathbb{C}^{d}\right)$ then
\begin{equation}
\underset{\ket S \sim \stab(4)}{\mathbb{E}} \ket S\!\bra S^{\otimes4}=\frac{1}{2^n\,D_{+}}\left(\Pi_{4}\Pi_{\mathrm{sym}}+\frac{4}{\left(d+4\right)}\left(I-\Pi_{4}\right)\Pi_{\mathrm{sym}}\right),
\end{equation}
where $D_{+}=\frac{\left(2^n+1\right)\left(2^n+2\right)}{6}=\trace(\Pi_{4}\Pi_{\mathrm{sym}})$.
\end{fact}

\section{Clifford testing via stabilizer testing}
\label{sec:reduction-clifford-to-stabilizer-testing}

In \cite[Remark 3.7]{grossSchurWeylDuality2021}, the authors observe that their 6-copy stabilizer testing algorithm also gives rise to a Clifford testing algorithm when applied to copies of the Choi state $\kett{U}$ of the unknown $n$-qubit unitary $U$. In this section, we formally establish this reduction from Clifford testing to stabilizer testing and extend it to the tolerant setting. This allows complexity guarantees demonstrated previously for stabilizer testing to be directly transferred to Clifford testing.
Since Choi states of Clifford unitaries are stabilizer states (\cref{lem:ClStab}), it follows that
\begin{equation}\label{eq:FClFStab}
    F_\mathrm{Cliff}(U) = \max_{C\in\mathrm{Cl}(n)}  |\langle\! \braket{C}{U}\!\rangle|^{2}\leq 
     \max_{S\in\stab(2n)}  |\braket{S}{U \rangle\!}|^2
     =
     F_{\stab}(\kett{U}).
\end{equation}
By definition, a stabilizer testing algorithm accepts a Choi state if it has high stabilizer fidelity.. In contrast, a Clifford testing algorithm should accept a unitary whenever it has high Clifford fidelity. 
But the relation~\cref{eq:FClFStab} does not exclude the possibility that the test is unsound.
To reduce Clifford testing to stabilizer testing, however, we also need a relation between the fidelities that goes in the opposite direction.

The main contributions of this section show that Clifford fidelity is polynomially related to the stabilizer fidelity of the Choi state and that equality even holds if the latter exceeds~1/2.

\begin{thm}[Equivalence of Clifford and stabilizer fidelity in high-fidelity regime]\label{thm:equivalence-fcliff-fstab}
    Let $U$ be an $n$-qubit unitary such that $F_{\stab}(\kett{U})> 1/2$. Then, 
    \begin{equation}
       F_\mathrm{Cliff}(U)= F_{\stab}(\kett{U}). 
    \end{equation}
\end{thm}

\begin{proof}
   We denote the first $n$ qubits as system $A$ and the second $n$ qubits as system $B$ so that
   $\kett{U}=(U\otimes I)\ket{\Omega}_{AB}$ where  $\ket{\Omega}_{AB}=\frac{1}{2^{n/2}} \sum_x \ket{x}_A \ket{x}_B$. 
    By the bipartite canonical form for stabilizer states \cite{fattalEntanglementStabilizerFormalism2004}, for all $\ket{S}\in \mathrm{Stab}(2n)$, there exists local Clifford unitaries $C_A,C_B$ and an integer number $r$ with $0\leq r\leq n$ such that 
    \begin{equation}
        \ket{S} = (C_A \otimes C_B) \big(\ket{\Phi}^{\otimes r} \otimes \ket{\sigma}\big) ,
    \end{equation}
    where $\ket{\Phi}$ denotes a $2$-qubit Bell state across the $(A|B)$-cut and $\ket{\sigma}$ is a product state on the remaining $2(n-r)$ qubits. 
    Hence, we have 
    \begin{equation}
        |\braket{S}{U \rangle\!}|^2 = \Big|\big( \bra{\Phi}^{\otimes r} \otimes \bra{\sigma}\big) \big|{U'\rangle\!}\rangle \Big|^2 \leq 2^{r-n},
    \end{equation}
    where $U' = C_A U C_B^T$, and the last inequality follows from using Schmidt-decompositions and Cauchy-Schwarz. Now, note that if $\ket{S}\not\in \{ \kett{C} \,|\; C\in \Cln \}$, then $r<n$. This proves the claim.
\end{proof}
\noindent \cref{thm:equivalence-fcliff-fstab} suffices to establish the Choi-state reduction for non-tolerant testing and thereby provides a rigorous underpinning of \cite[Remark 3.7]{grossSchurWeylDuality2021}.
\medskip

Next, we show that in general, the two fidelities can be quadratically far apart. Our proof is based on a probabilistic argument for which we require the following lemma:
\begin{lem}[Haar-random states typically have exponentially small stabilizer fidelity]\label{lem:levy-stab-states}
Let $\ket{\psi}$ be a Haar random $n$-qubit state. Then, for any constant $c<1$, it holds that
\begin{equation}
      \Pr_{\ket\psi}[ F_{\stab} (\ket{\psi}) \geq 2^{-cn} ] = \exp(- \Omega (2^{(1-c)n})) .
\end{equation}
\end{lem}

\begin{proof}
    By Levy's lemma (specifically the version given in \cite[Eq. 2]{gross2009most}), for a fixed state $\ket{\phi}\in \mathbb{C}^d$ and a Haar random state $\ket{\psi}\in\mathbb{C}^d$, we have
\begin{equation}
    \Pr_{\ket\psi} [|\braket{\phi}{\psi}|^2 \geq \epsilon] < \exp(- (2d-1) \epsilon). 
\end{equation}
The number of $n$-qubit stabilizer states is upper bounded as $|\stab_{n}|\leq 2^{\frac{1}{2}n^2 + 5n }$ (see \cite{AaronsonGottesman2004Improved}). Hence, by the union bound
\begin{equation}
     \Pr_{\ket\psi}[ F_{\stab} (\ket{\psi})\geq\eps ] \leq |\stab_{n}| \exp(- (2^{n+1}-1)\eps) .
\end{equation}
Choosing $\eps = 2^{-cn}$ with any constant $c<1$, the RHS is asympotically bounded as $\exp(- \Omega (2^{(1-c)n}))$.
\end{proof}

\begin{lem}[Inequivalence of Clifford and stabilizer fidelities]
    For sufficiently large $n$ and any $0\le k\le \f{n}{4}$,
        there exists an $n$-qubit unitary $U$, such that 
            $F_{\stab}\p{\kett{U}} \ge \f{1}{2^k}$ 
            and $F_{\cliff}(U) \le \f{1}{2^{2(k-1)}}$.
\end{lem}

\begin{proof}
    We define an $n$-qubit unitary $U$ as
    \begin{equation}
        U = \sum_{x\in \F^k} \ketbra{x}{x} \otimes U^{(x)},
    \end{equation}
    where $U^{(0)} = I$ and for all $x\in \F^k\setminus \lrs{0^k}$, 
        we pick $U^{(x)}$ independently Haar random over $n-k$ qubits.
    We will show that $F_{\stab}\p{\kett{U}} \ge \f{1}{2^k}$ whereas
        $\E_{U}\lrb{\sqrt{F_{\cliff}(U)}} \le \f{1}{2^{k-1}}$
            for sufficiently large $n$, then the desired conclusion follows.
    We have 
    \begin{equation}
        \kett{U} = \f{1}{\sqrt{2^n}}\p{\sum_{y\in \F^{n-k}} \ket{0^k y}\otimes\ket{0^k y}
            + \sum_{x\in \F^k\setminus \lrs{0^k}} \sum_{y\in \F^{n-k}}
                \ket{x}\otimes U^{(x)}\ket{y}\otimes \ket{xy}}.
    \end{equation}
    Consider the stabilizer state $\ket{S}:= \f{1}{\sqrt{2^{n-k}}} 
            \sum_{y\in \F^{n-k}} \ket{0^k y}\otimes \ket{0^k y}$. Its fidelity with the Choi state $\kett{U}$ gives the desired lower bound on the stabilizer fidelity:
    \begin{equation}
        F_{\stab}(\kett{ U})\ge \left|\braket{S}{U\rangle\!}\right|^{2}
            = \p{\f{1}{\sqrt{2^n}} \cdot \f{1}{2^{n-k}} \cdot 2^{n-k}}^2 
            = \f{1}{2^k}.
    \end{equation}
    On the other hand, we are going to show with high probability, 
        the fidelity between $U$ and any $n$-qubit Clifford unitary $C$ is small.
    For any $n$-qubit Clifford unitary $C$, we have
    \begin{align}
     |\langle\! \braket{C}{U}\!\rangle|
        &= \f{1}{2^{n}} \abs{\sum_{y\in \F^{n-k}} \bra{0^k\,y} C \ket{0^k\,y} 
            + \sum_{x\in \F^k\setminus\lrs{0^k}} \sum_{y\in \F^{n-k}} 
                \bra{xy} (I_{2^k}\otimes U^{(x)\dagger}) C\ket{xy}} \\
        &\le \frac{1}{2^{n}}\p{ 2^{\,n-k} + \sum_{x\in \F^k\setminus\lrs{0^k}} \sum_{y\in \F^{n-k}}
\big| \bra{\phi_{x,y}} U^{(x)}\ket{y} \big| },
    \end{align}
where $\ket{\phi_{x,y}} := (\bra{x}\otimes I) C^\dagger \ket{xy}$ is a (possibly sub-normalized) stabilizer state. Thus, for any $(n-k)$-qubit state $\ket{\psi}$, the overlap with $\ket{\phi_{x,y}}$ is bounded as 
\begin{equation}
    |\braket{\phi_{x,y}}{\psi}| \leq \|\phi_{x,y}\| \cdot \leq \max_{S\in \stab(n-k)} |\braket{S}{\phi_{x,y}}| \leq 
 \sqrt{F_{\stab}(\ket\psi)}.
\end{equation}
Hence, we have
\begin{equation}
     |\langle\! \braket{C}{U}\!\rangle| \leq  \f{1}{2^n} \p{2^{n-k} + \sum_{x\in \F^k\setminus\lrs{0^k}}
            \sum_{y\in \F^{n-k}} \sqrt{F_{\stab}\p{U^{(x)}\ket{y}}} }. 
\end{equation}
Now, we note that each $U^{(x)}\ket{y}$ is a Haar random $(n-k)$-qubit state, so by \cref{lem:levy-stab-states} and a union bound over all pairs $(x,y)$, we find
\begin{equation}
   \Pr \left[ \forall (x,y),\; F_{\stab} (U^{(x)}\ket{y}) \le 2^{-c(n-k)} \right] \leq 1- 2^n \cdot \exp(- \Omega (2^{(1-c)(n-k)})) = 1- o(1).
\end{equation}
Hence, choosing $c=0.98$ for $0\le k\le \f{n}{4}$, for sufficiently large $n$, we have
\begin{equation}
     |\langle\! \braket{C}{U}\!\rangle| \leq \f{1}{2^n}\p{2^{n-k} + 2^n \cdot 2^{-c(n-k)/2}} \leq 2^{-k} + 2^{-c(n-k)/2} \le \f{1}{2^{k-1}},
\end{equation}
with high probability. Since this bound holds uniformly over all $C\in \Cln$, the claim follows.
\end{proof}

To extend the reduction from to tolerant Clifford testing to tolerant stabilizer testing, we must relate the two fidelities also in the low-fidelity regime. The following general relation achieves this.

\begin{thm}[General relation between Clifford and stabilizer fidelity]\label{thm:relation-fstab-fcliff}
    Let $U$ be an $n$-qubit unitary. Then,
    \begin{equation}
  F_{\stab} (\kett{U})^6\leq  F_{\cliff}(U) \leq  F_{\stab}(\kett{U}).
    \end{equation}
\end{thm}
The upper bound is immediate from \cref{lem:ClStab}. To establish the lower bound, we develop a theory of Clifford testing parallel to that for stabilizer testing in \cite{grossSchurWeylDuality2021} and its subsequent extensions \cite{grewalImprovedStabilizerEstimation2024,arunachalamPolynomialTimeTolerantTesting2025,baoTolerantTestingStabilizer2025,mehrabanImprovedBoundsTesting2025}.

\subsection{The characteristic distribution of a unitary}

The central object in stabilizer testing is the characteristic distribution. 
In analogy, we will define a characteristic distribution for unitary operators via their Choi state as follows.

\begin{definition}[Characteristic distribution of a unitary]
Let $U$ be an $n$-qubit unitary. Then we define its corresponding characteristic distribution $p_U$ over $\mathbb{F}_2^{2n}\times \mathbb{F}_2^{2n}$ via its Choi state $\kett{U}$ as follows. For all $(x,y)\in \mathbb{F}_2^{2n}\times \mathbb{F}_2^{2n}$, let
    \begin{equation}
        p_U(x,y):=p_{\kett{U}}(x,y) = 2^{-2n}\left|\bbra{U}P_x\otimes P_y\kett{U} \right|^2.
    \end{equation}
\end{definition}
Note that we are abusing notation somewhat here by considering the pair $(x,y)=((a,b), (a',b'))$ in $p_{\kett{U}}(x,y)$ as an element of $\mathbb{F}_2^{4n}$ corresponding to the $2n$-qubit Pauli operator $P_x\otimes P_y$. The corresponding $X$ and $Z$ components are thus $(a,a')$ and  $(b,b')$, respectively.

We start by collecting some useful properties of the characteristic distribution $p_U$:
\begin{lem}[Properties of $p_U$]\label{lemma:propertiesofchardist}
Let $U$ be an $n$-qubit unitary. Then, the characteristic distribution has the following properties:
    \begin{enumerate}
        \item The probabilities $p_{U}(x,y)$ can be rewritten as
        \begin{equation}
            p_{U}(x,y) = \frac{1}{2^{4n}}\tr\left(P_{x}UP_{y}U^{\dagger}\right)^{2}.
        \end{equation}
        \item Marginalizing over $x$ or $y$ yields
        \begin{equation}
            \sum_{x\in \Ftn} p_{U}(x,y)= \sum_{y\in \Ftn} p_{U}(x,y)=2^{-2n}.
        \end{equation}
    \end{enumerate}
\end{lem}
\begin{proof}
    We note that $P_y = i^{a'\cdot b'}X^{a'}Z^{b'}$ and therefore
        $\overline{P_y} = i^{-a'\cdot b'}X^{a'}Z^{b'}$, because
        $\overline{X} = X$ and $\overline{Z} = Z$.
    We conclude that 
    \begin{equation}
     \frac{1}{2^{2n}}\left|\bbra{U}P_x\otimes P_y\kett{U} \right|^2 
        = \f{1}{2^{4n}} \abs{\tr \p{ P_x U \overline{P_y} U^\dagger}}^2
        = \f{1}{2^{4n}} \tr \p{ P_x U P_y U^\dagger}^2,
    \end{equation}
    where the last equality follows from the fact that $\tr \p{ P_x U P_y U^\dagger}$ is a real number.
    To prove the marginalization property, fix any $y$. Then, we have
    \begin{equation}
        \sum_{x} p_U(x,y)
          = \f{1}{2^{4n}} \sum_x \abs{\tr \p{ P_x U P_y U^\dagger}}^2
          = \f{1}{2^{3n}} \norm{UP_yU^\dagger}_{2}^2 = 2^{-2n},
    \end{equation}
    where we used Parselval's identity from \cref{eq:parseval}.
    A analogous argument holds for any fixed $x$ and summing over $y$.
\end{proof}

\begin{lem}[Bound on collision probability]\label{lemma:Q2inverse}
    Let $U$ be a unitary on $n$ qubits. It holds that
\begin{equation}
        \max_{x\in \Ftn} \abs{\widehat{U}(x)}^2 \geq \sum_{x\in \Ftn} \abs{\widehat{U}(x)}^4= \sum_{x\in \Ftn} p_{U}(x,x).
    \end{equation}
\end{lem}
\begin{proof}
     Using the Fourier inversion formula~\cref{eq:fourier_inversion},  we have 
    \begin{align}\label{eq:diagonal-sum}
        \sum_{x\in \Ftn}p_U(x,x) &= \f{1}{2^{4n}}\sum_{x\in \Ftn}\tr\p{P_xUP_xU^\dagger}^2,\\
        \nonumber
        &= \sum_{x,y_1,y_2\in \Ftn} \f{1}{2^{4n}}\abs{\widehat{U}(y_1)\widehat{U}(y_2)}^2
            \tr\p{P_x P_{y_1} P_x P_{y_2}}^2, \\
             \nonumber
        &= \sum_{y_1,y_2\in \Ftn} \f{1}{2^{2n}}\abs{\widehat{U}(y_1)\widehat{U}(y_2)}^2
            \tr\p{P_{y_1}P_{y_2}}^2 ,\\
             \nonumber
        &= \sum_{y\in \Ftn} \abs{\widehat{U}(y)}^4, \\
         \nonumber
        &\le \max_{x} \abs{\widehat{U}(x)}^2 \sum_{y\in \Ftn} \abs{\widehat{U}(y)}^2 ,\\
         \nonumber
        &= \max_{x} \abs{\widehat{U}(x)}^2,
         \nonumber
    \end{align}
    where the last equality follows from Parseval's identity~\cref{eq:parseval}.
\end{proof}

\subsection{Clifford Lagrangians}
A key result in stabilizer testing is \cref{fact:stab-fidelity-lower-bound}, showing that the stabilizer fidelity of a state~$\ket{\psi}$ is in general bounded from below by the weight of any Lagrangian subspace under the characteristic distribution of~$\ket{\psi}$.
For Clifford testing, we would like to establish an analogous inequality using the Clifford fidelity $F_{\mathrm{Cliff}}(U)$ and $p_U$ instead. However, to do so, we have to restrict our attention to a subset of Lagrangian subspaces of $\Ftn \times \Ftn$ that is in correspondence to Clifford Choi states.

Stabilizer groups (upon forgetting phases) are in 1-to-1 correspondence to Lagrangian subspaces. Similarly, the stabilizer groups of Clifford Choi states are in 1-to-1 correspondence to Clifford Lagrangian subspaces.

\begin{definition}[Clifford Lagrangian subspace]
    A Lagrangian subspace $M\subset \Ftn \times \Ftn$ is a called a \textit{Clifford Lagrangian subspace} if there exists $S\in \mathrm{Sp}(2n,\F)$ such that $M$ is the graph of $S$:
    \begin{equation}
        M= \{(x,Sx): x\in \Ftn \}.
    \end{equation}
\end{definition}

With this, we can now prove an inequality analogous to \cref{fact:stab-fidelity-lower-bound}:

\begin{fact}[Lower bound for the Clifford fidelity]\label{fact:cliff-fidelity-lower-bound}
Let $U$ be an $n$-qubit unitary and let $M\subset \Ftn \times \Ftn$ be a Clifford Lagrangian subspace. Then,
    \begin{equation}
        F_{\mathrm{Cliff}}(U) \geq p_U(M).
    \end{equation}
\end{fact}
\begin{proof}
Since $M$ is a Clifford Lagrangian subspace, there exists $S\in \mathrm{Sp}(2n,\mathbb{F}_2)$ such that $M = \{(x,Sx): x\in \Ftn \}$.
This implies the existence of a Clifford $C$ such that for all $x\in \F^{2n}$ we have
\begin{equation}
    CP_{Sx}C^\dagger = \pm P_{x}.
\end{equation}
\noindent Hence, the weight of $p_U$ on the Lagrangian subpace $V$ can be expressed as
\begin{align}
    p_U(M)=\sum_{x\in \F^{2n}}p_{U}(x,Sx)
    =&\sum_{x}\frac{1}{2^{4n}}\tr\left(P_xUP_{Sx}U^\dagger\right)^2\\
    \nonumber
    =&\sum_{x}\frac{1}{2^{4n}}\tr\left(P_xUCP_{x}C^\dagger U^\dagger\right)^2\\
     \nonumber
    =&\sum_{x}p_{UC}(x,x).
     \nonumber
\end{align}
By \cref{lemma:Q2inverse}, there exists a Pauli $P_x$ such that 
\begin{equation}
\abs{\frac{1}{2^{n}}\tr\p{P_x^\dagger U C}}^2 = \abs{\frac{1}{2^{n}}\tr\p{\p{P_x C^\dagger}^\dagger U}}^2\geq \sum_{x}p_{UC}(x,x).
\end{equation}
Since $P_x C^\dagger$ is a Clifford, we have that 
\begin{equation}
    F_{\cliff}(U)\geq \sum_{x}p_{UC}(x,x)=p_U(M).
\end{equation}
\end{proof}

While $F_{\cliff}(U)$ is naturally bounded in terms of Clifford Lagrangian subspaces, $F_{\stab}(\kett{U})$ is related to arbitrary Lagrangian subspaces. To relate these two notions, we analyze how the characteristic distribution $p_U$ behaves on isotropic subspaces of $\Ftn\times\Ftn$. In particular, we show that every isotropic subspace contains a large-weight component that can be extended to a Clifford Lagrangian subspace. Intuitively, this component is obtained by removing the degenerate parts of the subspace that prevent it from being a graph of a symplectic map.

\begin{definition}[Extendability to a Clifford Lagrangian]
Let $V\subset \Ftn \times \Ftn$ be an isotropic subspace. We say that $V$ is \textit{extendable to a Clifford Lagrangian subspace} if there exists $S\in \mathrm{Sp}(2n,\F)$ such that 
\begin{equation}
    V \subset \{(x,Sx): x\in \Ftn \}.
\end{equation}
\end{definition}

\begin{lem}[Every isotropic subspace contains a subspace that is extendable.]\label{lem:characterization-extendability}
    Let $V\subset \Ftn \times \Ftn$ be an isotropic subspace. Let
    \begin{equation}
        L_0 =\{x\in \Ftn:(x,0)\in V \}, \quad R_0 =\{y\in \Ftn:(0,y)\in V \},
    \end{equation}
    and let $V'\subseteq V$ be such that
    \begin{equation}
        V = V' \oplus (L_0 \oplus 0) \oplus (0 \oplus R_0).
    \end{equation}
    Then $V'$ is extendable to a Clifford Lagrangian subspace.   
\end{lem}
\begin{proof}
Let $\pi_L : \Ftn \times \Ftn \to \Ftn$ and $\pi_R : \Ftn \times \Ftn \to \Ftn$ denote the projections onto the first and second coordinates, respectively, and set
\begin{equation}
    L' = \pi_L(V'), \quad R' = \pi_R(V'),
\end{equation}
with $L',R'\subseteq \Ftn$.

Then $V'$ is the graph of a bijective linear map $F: L'\to R'$, meaning that 
\begin{equation}
    V' = \{(x,F x): x\in L'\}.
\end{equation}

\noindent Furthermore, since $V$ is isotropic we have
\begin{equation}
    [(x_1,y_1),(x_2,y_2)] = [x_1,x_2]+[y_1,y_2]= 0, \quad \forall\, (x_1,y_1), (x_2,y_2) \in V',
\end{equation}
which is equivalent to
\begin{align}
    [x_1,x_2]=[Fx_1,Fx_2], \quad \forall x_1,x_2 \in L'.
\end{align}
 This means $F$ also preserves the symplectic inner product and is hence a symplectic isometry between the subspaces $L'$ and $R'$. By Witt's extension theorem \cite[Theorem 3.8]{artin2016geometric}, every symplectic isometry between subspaces extends to a global symplectic automorphism on whole space. Hence, there exists $S\in \mathrm{Sp}(2n, \mathbb{F}_2)$ extending $F$. Therefore, $V' \subseteq \{(x,Sx): x\in \Ftn \}$, as required.
\end{proof}

\begin{lem}[High weight extendable subspace]\label{lem:weight-of-extendable-subspace}
Let $V$ be an isotropic subspace of~$\Ftn \times \Ftn$. Then there exists a subspace $V'\subseteq V$ such that $V'$ is extendable to a Clifford Lagrangian and
    \begin{equation}
        p_U(V') \geq p_U(V)^3.
    \end{equation}
\end{lem}
\begin{proof}
Let $V \subseteq \Ftn \times \Ftn$ be an isotropic subspace, and let $V',L_0,R_0$ be as in \cref{lem:characterization-extendability}. Then $V'$ is extendable to a Clifford Lagrangian.

We now show that $V'$ has high weight.
By the Pigeonhole principle, there exist $x_0\in L_0$ and $y_0\in R_0$ such that
\begin{align}\label{eq:pigeonhole}
    p_{U
    }(V'+(x_0,y_0))\geq \frac{1}{|L_0|\cdot |R_0|}p_{U
    }(V).
\end{align}  

By \Cref{lem:removeshifts}, $V'$ itself also has high weight: 
\begin{align}\label{eq:removeshifts}
    p_{U}(V')\geq p_{U}(V'+(x_0,y_0)).
\end{align}
What is left to show is that $\frac{1}{|L_0|}$ and $\frac{1}{|R_0|}$ are greater than $p_U(V)$. We do this next.

Let $L=\pi_L(V)$ and $R=\pi_R(V)$ where $\pi_L,\pi_R$ are as in \cref{lem:characterization-extendability}.
By the rank-nullity theorem,
\begin{equation}\label{eq:dimensionsleftrightkernels}
    \dim L+\dim R_0=\dim L_0+\dim R =\dim V
\end{equation}
The weight of $V$ can be upper bounded as
\begin{align}
    p_{U}(V)
    =& \sum_{(x,y)\in V}p_U(x,y)
        \nonumber\\
    \nonumber
    \leq&\sum_{x\in L}\sum_{y\in \F^{2n}}p_U(x,y)\\
    =&|L|2^{-2n},\label{eq:sizeL}
\end{align}
where we have used \Cref{lemma:propertiesofchardist} in the last equality.

Since $V$ is isotropic, $|V|\leq 2^{2n}$. Using this and \Cref{eq:dimensionsleftrightkernels}, we can upper bound \Cref{eq:sizeL} by 
\begin{align}
    |L|2^{-2n}\leq \f{|V|}{|R_0|}2^{-2n}\leq \f{1}{|R_0|},
\end{align}
and therefore 
\begin{align}
\frac{1}{|R_0|}\geq p_{U
    }(V).\label{eq:R0issmall}
\end{align}
In a similar fashion, we can prove that
\begin{align}
    \frac{1}{|L_0|}\geq p_{U
    }(V).\label{eq:L0issmall}
\end{align}

Combining \Cref{eq:pigeonhole}, \Cref{eq:removeshifts}, \Cref{eq:R0issmall} and \Cref{eq:L0issmall}, we have that 
\begin{align}
    p_{U}(V')\geq p_{U}(V)^3.
\end{align}
\end{proof}

\subsection{Relating Clifford fidelity to stabilizer fidelity}

We now have established all the ingredients to prove that Clifford fidelity is polynomially related to the stabilizer fidelity of the Choi state.

\begin{proof}[Proof of \cref{thm:relation-fstab-fcliff}]
The inequality $F_{\stab}(\kett{U})\geq  F_{\cliff}(U)$ follows directly from \cref{lem:ClStab}.

To prove that $F_{\cliff}(U) \geq F_{\stab}(\kett{U})^6$ we apply \cref{lem:upperbound-stabilizer-fidelity} to the Choi state $\kett{U}$ to find:
\begin{equation}\label{eq:upper-bound-on-choi-state-stab-fidelity}
     \max_{M \text{ Lagrangian}}p_{U}(M) \ge F_{\stab}(\kett{U})^2.
\end{equation}
Let $M^*$ be the Lagrangian subspace attaining the maximum in \cref{eq:upper-bound-on-choi-state-stab-fidelity}. By \cref{lem:weight-of-extendable-subspace}, there is a subspace $V'\subset M^*$ that satisfies
    \begin{equation}
              p_U(V') \geq p_U(M^*)^3.
    \end{equation}
    Lastly, since $V'$ is extendable, there exists a Clifford Lagrangian subspace $M'$ such that $V'\subseteq M'$ and by \cref{fact:cliff-fidelity-lower-bound}, $F_{\cliff}(U)\geq p_U(M') $. 
    Combining these inequalities yields,
    \begin{equation}
        F_{\cliff}(U)\geq p_U(M') \geq p_U(V')\geq p_U(M^*)^3 \geq F_{\stab}(\kett{U})^6.
    \end{equation}
\end{proof}

\section{A 4-query Clifford tester}\label{sec:4-query-tester}
In this section we will present our 4-query Clifford testing algorithm.

To build intuition for our algorithm, it is helpful to contrast the action of the Clifford group on multiple copies. For $t\leq 3$, the action of the Clifford group on $t$ copies is indistinguishable from that of the full unitary group since the commutants coincide. At $t=4$, however, the situation changes:
Ref.~\cite{zhuCliffordGroupFails2016} first showed that there exists a subspace $V_{n,4}\subset ((\mathbb{C}^2)^{\otimes n})^{\otimes 4}$ that is invariant under the diagonal Clifford action $C^{\otimes 4}$ for all $C\in \Cln$, but is not invariant under the corresponding $t=4$-fold Haar group twirl of the unitary group.
The projector onto this Clifford-invariant subspace $V_{n,4}$ is denoted $\Pi_4$ (see \cref{eq:pi_4_definition}).

This observation suggests a natural 4-query test to distinguish Clifford unitaries from Haar-random unitaries. The test works as follows:
\begin{enumerate}
    \item Prepare the uniform mixture over $V_{n,4}$, i.e. prepare the mixed state $\rho = \Pi_4 / \tr(\Pi_4)$.
    \item Apply $U^{\otimes 4}$.
    \item Measure the projection onto $V_{n,4}$, i.e. measure the two-outcome POVM $\{\Pi_{4}, I-\Pi_{4}\}$.
\end{enumerate}
Intuitively, this procedure checks if the subspace $V_{n,4}$ is left invariant under the action of the unitary.

This is precisely the test we propose for Clifford testing. Our contribution is to analyze this test in detail: we show that it not only separates Cliffords from Haar random unitaries, but also distinguishes Clifford from non-Clifford unitaries up to any desired $\epsilon$ distance. Hence, it constitutes a Clifford testing algorithm. Moreover, we show that this test is also a tolerant test.

We stress, however, that although a Clifford-invariant subspace already exists at $t=4$, this does not yield a 4-copy tester for stabilizer states: in fact, stabilizer testing is known to require at least $t=6$ copies \cite{damanikMScThesisAfstudeerscriptie,grossSchurWeylDuality2021, grevinkWillItGlue2025}. Our 4-query tester is therefore genuinely specific to Clifford testing, and not in conflict with the known lower bounds for stabilizer testing.

We start our exposition by presenting a more space-efficient implementation of the above test. In particular, while a naive implementation of this process would require $4n$ qubits of workspace, our implementation in \cref{alg:4cltester} below only uses $2n$ qubits of space. 
The key observation here is that $V_{n,4}$ admits a basis that factorizes into tensor products of $2n$-qubit Bell states (see \cref{eq:basis}), so that 
\begin{equation}\label{eq:Pi4_expansion}
    \Pi_4 = \sum_{x\in\Ftn} (\kett{P_x}\bbra{P_x})^{\otimes 2}.
\end{equation}
\noindent Moreover, we emphasize that this 4-query algorithm is also more space-efficient than the Choi-state-based reduction from the 6-copy stabilizer tester discussed in \cref{sec:reduction-clifford-to-stabilizer-testing}, which uses at least $4n$ qubits of workspace to perform Bell difference sampling.

Our algorithm proceeds as follows:
\medskip

\begin{algorithm}[H]
\caption{Four-query Clifford tester}
\label{alg:4cltester}
\DontPrintSemicolon
\SetKwInOut{Input}{Input}
\Input{Black-box access to an $n$-qubit unitary $U$.}
\medskip
$x \gets \mathrm{Uniform}(\Ftn)$\tcp*{sample a random label}
Prepare two independent copies of $U^{\otimes 2}\kett{P_x}$\;
Measure each copy in the Bell basis $\{\kett{P_y}\bbra{P_y}\}_y$ to obtain outcomes $y$ and $y'$\;
\lIf{$y = y'$}{\Return{\textbf{Accept}}}
\lElse{\Return{\textbf{Reject}}}

\medskip
\textbf{Queries to $U$:} $4$.
\end{algorithm}
\medskip

Let us now turn to analyzing \cref{alg:4cltester}:
By \cref{eq:Pi4_expansion}, the acceptance probability of the 4-query test is given by
\begin{equation}\label{eq:pacc_pi_4}
p_{\rm acc}\left(U\right)=\frac{1}{2^{2n}}\sum_{x\in \Ftn} \trace\left(U^{\otimes 4} (\kett{P_x}\bbra{P_x})^{\otimes 2} U^{\dagger}{}^{\otimes4} \Pi_4 \right)=\frac{1}{2^{2n}}\trace\left(\Pi_{4}U^{\otimes4}\Pi_{4}U^{\dagger}{}^{\otimes4}\right).
\end{equation}
Using that $\Pi_4=2^{-2n} \sum_{x}P_{x}^{\otimes4}$ from \cref{eq:pi_4_definition} and that $p_{U}\left(x,y\right)=2^{-4n}\trace\left(P_{x}UP_{y}U^{\dagger}\right)^{2}$ from \cref{lemma:propertiesofchardist}, the acceptance probability can be rewritten in terms of the characteristic distribution $p_U$ of the unitary $U$ as 
\begin{align}
p_{\rm acc}\left(U\right) 
& =\frac{1}{2^{6n}}\sum_{x,y\in \Ftn}\trace\left(P_{x}UP_{y}U^{\dagger}\right)^{4}
\nonumber
\\
 & =2^{2n}\sum_{x,y\in \Ftn}p_{U}\left(x,y\right)^{2} \label{eq:pacc-pU}\\
 & =2^{2n}\left\Vert p_{U}\right\Vert _{2}^{2}. \label{eq:pU-2-norm}
\end{align}
\begin{remark}The appearance of the $2$-norm of the characteristic distribution $p_U$ is reminiscent of the $6$-copy stabilizer testing algorithm from 
Ref.\ \cite[Eq. (3.14)]{grossSchurWeylDuality2021} based on Bell difference sampling whose acceptance probability $p_{\rm acc}^{\text{GNW}}\left(\ket{\psi}\right)$ features the $3$-norm of the characteristic distribution $p_{\ket{\psi}}$ of the $n$-qubit state $\ket{\psi}$, 
\begin{equation}
p_{\rm acc}^{\text{GNW}}\left(\ket{\psi}\right)=\frac{1}{2}\left(1+2^{2n}\left\Vert p_{\ket{\psi}}\right\Vert _{3}^{3}\right).
\end{equation}
Applying this $6$-copy tester directly to the Choi state $\kett{U}$ of the unknown $n$-qubit unitary $U$, it would accept with probability
\begin{equation}
p_{\rm acc}^{\text{GNW}}\left(\kett{U}\right)=\frac{1}{2}\left(1+2^{4n}\left\Vert p_U\right\Vert _{3}^{3}\right).
\end{equation}
\end{remark}

Next, we relate the acceptance probability to the stabilizer fidelity of the Choi state $\kett{U}$:
\begin{lem}[Bound on acceptance probability]
\label{lem:bound-on-acceptance-probability}
Let $U$ be an $n$-qubit unitary. Then, the acceptance probability of \cref{alg:4cltester} is upper bounded as follows
    \begin{equation}
        p_{\rm acc}\left(U\right) = 2^{2n}\left\Vert p_{U}\right\Vert _{2}^{2}\leq \frac{1+F_{\stab}(\kett{U})}{2}  .
    \end{equation}
\end{lem}
\begin{proof}
Define the set
\begin{equation}
M_{U}:=\left\{ \left(x,y\right)\in\mathbb{F}_{2}^{2n}\times\mathbb{F}_{2}^{2n}:2^{2n}p_{U}\left(x,y\right)>1/2\right\} .
\end{equation}
By \cref{fact:uncertainty-principle}, $M_U$ can be extended to a Lagrangian subspace of $\mathbb{F}_{2}^{2n}\times\mathbb{F}_{2}^{2n}$. 
Using \cref{fact:stab-fidelity-lower-bound} on the Choi state $\kett{U}$, we have
\begin{equation}\label{eq:inequality-1}
F_{\mathrm{Stab}} \left(\kett{U}\right)\geq p_U(M) \geq p_U(M_U)
\end{equation}

\noindent We will now show
\begin{equation}\label{eq:inequality-2}
p_U(M_U)\geq 2\cdot 2^{2n}\left\Vert p_{U}\right\Vert _{2}^{2}-1.
\end{equation}

Recall that $ p_U(M_U) = \sum_{(x,y)\in M_{U}} p_{U}\left(x,y\right)$. Using Markov's inequality, we find
\begin{align}
\sum_{\left(x,y\right)\in M_{U}}p_{U}\left(x,y\right) & =\Pr_{\left(x,y\right)\sim p_{U}}\left[p_{U}\left(x,y\right)\in M_{U}\right]\\
\nonumber
 & =\Pr_{\left(x,y\right)\sim p_{U}}\left[2^{2n}p_{U}\left(x,y\right)>1/2\right]\\
 \nonumber
 & =1-\Pr_{\left(x,y\right)\sim p_{U}}\left[1-2^{2n}p_{U}\left(x,y\right)\geq1/2\right]\\
 \nonumber
 & \geq1-2\left(\mathbb{E}_{\left(x,y\right)\sim p_{U}}\left[1-2^{2n}p_{U}\left(x,y\right)\right]\right)\\
 \nonumber
 & =2^{2n+1}\mathbb{E}_{\left(x,y\right)\sim p_{U}}\left[p_{U}\left(x,y\right)\right]-1\\
 \nonumber
 & =2^{2n+1}\left\Vert p_{U}\right\Vert _{2}^{2}-1.\\
\end{align}
Combining \cref{eq:inequality-1,eq:inequality-2} yields the claimed relation.
\end{proof}

\subsection{One-sided Clifford testing}

We now have all established all ingredients to show that \cref{alg:4cltester} constitutes a Clifford tester.

\fourtest*

\begin{proof}
\noindent We consider completeness and soundness of the test separately:\\

Perfect completeness follows immediately since, if $U$ is a Clifford
unitary, then $\left[U^{\otimes4},\Pi_{4}\right]=0$ and hence $p_{\rm acc}\left(U\right)=1$ by \cref{eq:pacc_pi_4}.\\

On the other hand, for the soundness analysis, assume $F_{\mathrm{Cliff}}\left(U\right)\leq 1-\varepsilon$. By \cref{lem:bound-on-acceptance-probability}, we have that $
    p_{\rm acc}\left(U\right) \leq \frac{1}{2}\left(1+F_{\stab}(\kett{U})\right)$.
Now, we distinguish two cases: If 
$F_{\mathrm{Cliff}}\left(U\right)\leq1/2$, then $
p_{\rm acc}\left(U\right)\leq3/4$. On the other hand, if 
$F_{\mathrm{Cliff}}\left(U\right)>1/2$, then by \cref{thm:equivalence-fcliff-fstab}, $F_{\mathrm{Cliff}}\left(U\right)=F_{\stab}(\kett{U})$ and so  $p_{\rm acc}\left(U\right)\leq 1-\frac{\varepsilon}{2}$ which completes the proof.
\end{proof}

By repeating the test $O\big(1/\varepsilon)$ times and rejecting if any single run rejects, we can boost the soundness case to an arbitrary success probability. 
This is formalized in the following corollary.

\begin{cor}
For any $\eps>0$, there is an $\eps$-Clifford tester that makes $O(1/\eps)$ queries.
\end{cor}

\subsection{Tolerant Clifford testing}\label{ssec:tolerant-4-query-test}

Next, we show that \cref{alg:4cltester} is a tolerant tester.
To this end, we make a connection to the Gowers uniformity norms as well as the quantum uniformity measures introduced in Ref.\ \cite{buQuantumHighOrderAnalysis2025}.
The analysis proceeds roughly in four steps.
First show that, on input~$U$, the acceptance probability of our test is proportional to the quantum uniformity measure of~$U$.
Second, we show that in turn, this equals the Gowers $U^3$ norm of the Choi state of~$U$.
Third, we use an inverse theorem for the~$U^3$ norm showing that it is polynomially equivalent to the stabilizer fidelity.
Fourth, we use our polynomial relation between Clifford and stabilizer fidelity from \cref{thm:relation-fstab-fcliff}.
\medskip

We begin by recalling the definition of the Gowers uniformity norms.

\begin{definition}[Gowers uniformity norms]
    For a function~$f:\F^n\to \mathbb{C}$ and $h\in \F^n$, define the multiplicative derivative of~$f$ in direction~$h$ to be the function given by $\Delta_hf(x) = f(x+h)\overline{f(x)}$.
    For every natural number~$k\geq 1$, the Gowers $U^k$ norm of~$f$ is then given by
    \begin{equation}
        \|f\|_{U^k} = \Big(\sum_{x,h_1,\dots,h_k\in \F^n}\Delta_{h_k}\cdots\Delta_{h_1}f(x)\Big)^{\frac{1}{2^k}}.
    \end{equation}
For an $n$-qubit pure state~$\ket{\psi}$, we let $\|\ket{\psi}\|_{U^k}$ denote the $U^k$ norm of the function giving its amplitudes in the computational basis.
\end{definition}

\begin{remark}
    In the literature, the uniformity norms are usually defined using expectations instead of sums.
    We break with this tradition to avoid dimension factors appearing due to the fact that in quantum computing, Hilbert spaces are usually defined using the counting measure (as opposed to the uniform probability measure).
    The only difference is of course nothing more than a rescaling.
\end{remark}

We will use the following lemma from~\cite[Lemma~3.3]{arunachalamPolynomialTimeTolerantTesting2025}.

\begin{lem}[Arunachalam and Dutt]\label{lem:U3char}
    For any $n$-qubit quantum state~$\psi$, we have that
    \begin{equation}
        \|\ket{\psi}\|_{U^3}^8 = 2^{n}\|p_{\ket{\psi}}\|_2^2.
    \end{equation}
\end{lem}

Furthermore, we will use an inverse theorem for the $U^3$ norm of pure states, a result that was obtained roughly concurrently but independently in~\cite{baoTolerantTestingStabilizer2025, ArunachalamBravyiDutt2024, mehrabanImprovedBoundsTesting2025}.

\begin{thm}[Inverse theorem for $U^3$ norm of quantum states] \label{thm:state-inverse-theorem}
Let $\ket\psi$ be an $n$-qubit quantum state. Then, 
\begin{equation}
    F_{\stab}(\ket{\psi}) \geq \poly\big(\|\ket{\psi}\|_{U^3}\big).
\end{equation}
\end{thm}

\begin{definition}[Quantum uniformity measures]\label{def:uniformity-norm}
    For a matrix $A\in\C^{\F^n\times\F^n}$ and $x\in\F^n\times\F^n$, define the multiplicative derivative of~$U$ in direction~$x$ to be the matrix given by $\partial_x A = P_x U P_x^\dagger U^\dagger$.
    For every natural number $k\ge 1$, the $Q^k$ norm of~$U$ is then given by
    \begin{equation}
        \norm{A}_{Q^k} = \Big(\displaystyle\mathop{\E}_{x_1,\dots,x_k\in \Ftn} 
            \f{1}{2^n}\tr\lrb{\partial_{x_k}\cdots\partial_{x_1} A}\Big)^{\frac{1}{2^k}}.
    \end{equation}
    
\end{definition}

The following lemma shows that \cref{lem:U3char} generalizes to the non-commutative setting.

\begin{lem}\label{lem:QpU}
    For any $n$-qubit unitary~$U$, we have that
    \begin{align}
         \norm{U}_{Q^2}^4 &= \sum_{x \in \Ftn}p_U(x,x)\label{eq:fourth-power}\\
        \norm{U}_{Q^3}^8 &=  2^{2n}\lVert p_U \rVert_2^2.\label{eq:Q3pU}
    \end{align}
\end{lem}
         
\begin{proof}
We use the following two elementary properties of the $Q^k$ norms~\cite{buQuantumHighOrderAnalysis2025}.
First, the $Q^1$ norm is in fact a semi-norm:
\begin{equation}
    \|A\|_{Q^1} = \big|\tfrac{1}{2^n}\tr(A)|.
\end{equation}
Second, we have the nesting property:
\begin{equation}
    \norm{A}_{Q^k}^{2^k}  = \E_{x\in \Ftn} \norm{\partial_x A}_{Q^{k-1}}^{2^{k-1}}.
\end{equation}
It follows from these identities that
    \begin{equation}
        \norm{U}_{Q^2}^4 =  \underset{x\in \Ftn}{\mathbb{E}} \norm{\partial_x U}_{Q^1}^2
            = \underset{x\in \Ftn}{\mathbb{E}}\abs{\f{1}{2^n}\tr\lrb{\partial_x U}}^2 = \frac{1}{2^{4n}}
             \sum_{x\in \Ftn}  \tr\lrb{P_x U P_x U^\dagger}^2 =
            \sum_{x \in \Ftn}p_U(x,x),
    \end{equation}
which proves~\Cref{eq:fourth-power}.
  Combining this with~\cref{lemma:Q2inverse} also gives 
      \begin{align}
        \norm{U}_{Q^3}^8 &=  \underset{x\in \Ftn}{\mathbb{E}}  \norm{\partial_x U}_{Q^2}^4
          =   \frac{1}{2^{2n}} \sum_{x,y\in \Ftn}\abs{\widehat{\partial_xU}(y)}^4
          = \frac{1}{2^{2n}} \abs{\f{1}{2^n} \sum_{x,y}\tr\lrb{P_y P_x U P_x^\dagger U^\dagger}}^4 ,\\
          &=  2^{2n} \sum_{x,y} p_U(x+y,x)= 2^{2n} \sum_{x,y} p_U(x,y)^2.
    \end{align}
This proves~\Cref{eq:Q3pU}.
\end{proof}

Combining \cref{lem:QpU} and \cref{lem:bound-on-acceptance-probability}  shows that the acceptance probability of \cref{alg:4cltester} is equal to the eighth power of the $Q^3$ norm. 
Moreover, we get that the $Q^3$ norm can be written as the $U^3$ norm of the Choi state.

\begin{cor}\label{cor:Q3U3}
    Let~$U$ be an $n$-qubit unitary operator. Then,
    \begin{equation}
        \|U\|_{Q^3} = \|\kett{U}\|_{U^3}.
    \end{equation}
\end{cor}

\begin{proof}
    Since it trivially holds that $\|p_U\|_2 = \|p_{\kett{U}}||_2$, \cref{lem:U3char} and \cref{lem:QpU} then give that
\begin{align}
\|\kett{U}\|_{U^3}^8 &= 2^{2n}\|p_{\kett{U}}\|_2^2\\
    &=2^{2n}\|p_U\|_2^2\\
    &=\|U\|_{Q^3}^8. 
\end{align}
This proves the claim.

\end{proof}

From this, we now easily obtain inverse theorem for the~$Q^3$ norm, which resolves~\cite[Conjecture~1]{buQuantumHighOrderAnalysis2025} and may be of independent interest.

\Qinverse*

\begin{proof}
\cref{thm:relation-fstab-fcliff} and \cref{thm:state-inverse-theorem} immediately imply that $F_{\mathrm{Cliff}}(U) \geq \poly\big(\|\kett{U}\|_{U^3}\big)$.
The result now follows from \cref{cor:Q3U3}.
\end{proof}

In turn, it follows that the 4-query tester in \cref{alg:4cltester} constitutes a tolerant tester:

\tolerantfourtest*

\begin{proof}
    Consider again the acceptance probability of the $4$-query test. From \Cref{eq:pacc-pU} we have
    \begin{equation}
    p_{\rm acc}(U) = 2^{2n}\lVert p_U \rVert_2^2.
    \end{equation}

Applying \cref{lem:upperbound-stabilizer-fidelity} to the Choi state $\kett{U}$, then completeness now follows from \Cref{cor:upperbound-fidelity} in the following way. If $F_{\mathrm{Cliff}}(U) \ge \varepsilon$, we have 
    \begin{equation}
        p_{\rm acc}(U) = 2^{2n}\lVert p_U \rVert_2^2  
        =
        2^{2n}\lVert p_{\kett{U}} \rVert_2^2
        \geq F_{\mathrm{Stab}}(\kett{U})^4 \ge F_{\cliff}(U)^4 \geq \varepsilon^4.
    \end{equation}
 Soundness follows immediately from \cref{lem:QpU} and \cref{thm:Qinverse}, since together, they give
\begin{equation}
    p_{\rm acc}(U)
    =
    \|U\|_{Q^3}^8
    \leq
    \poly(F_{\mathrm{Cliff}}(U)).
\end{equation}
This proves the result.
\end{proof}

\begin{cor}
    For any $\eps>0$, there exists a $\poly(\eps)$-query $(c^{-1}\eps^c,\eps)$-tolerant Clifford tester, where $c\geq 1$ is an absolute constant.
\end{cor}

\begin{remark}
    Close inspection of the proof shows that one may take $c = 2688$.
\end{remark}

\section{Auxiliary-free single-copy Clifford testing}\label{sec:single-copy-upper-bound}

In this section, we give a single-copy Clifford testing algorithm, by which we mean that the algorithm, immediately after each query to~$U$, applies it to some input state of our choice and measures
before querying $U$ again. 
It does not keep any coherent quantum memory register between such prepare-and-measure rounds. Importantly, our algorithm is  auxiliary-free, meaning it does not require any extra auxiliary qubits apart from the $n$-qubit register that the unknown unitary $U$ acts on.
Our algorithm builds on the single-copy stabilizer testing algorithm given in Ref.\ \cite{hinscheSingleCopyStabilizerTesting2025}.

\begin{thm}[Single-copy stabilizer testing algorithm from Ref.~\cite{hinscheSingleCopyStabilizerTesting2025}]\label{thm:guarantees-stab-tester}
    There exists a single-copy stabilizer testing algorithm that, given parameters $\eps,\delta>0$ and $O\big(\frac{n}{\varepsilon^2} \log \frac{1}{\delta}\big)$ copies of an unknown state~$\ket{\psi}$, has the following guarantees:
    \begin{itemize}
        \item It accepts with probability at least~$1 - \delta$ if $\ket{\psi}$ is a stabilizer state.
        \item It rejects with probability at least $1 - \delta$ if $F_{\mathrm{Stab}}\left(\ket{\psi} \right)\leq 1-\varepsilon$.
    \end{itemize}
    Moreover, the algorithm runs in time~$O\big(\frac{n^3}{\varepsilon^2} \log \frac{1}{\delta}\big)$.
\end{thm}

Based on this, our auxiliary-free single-copy Clifford tester then proceeds as follows:
\medskip

\begin{algorithm}[H]
\caption{Auxiliary-free single-copy Clifford tester}
\label{alg:aux-free-single-copy-tester}
\DontPrintSemicolon
\SetKwInOut{Input}{Input}
\Input{Parameter $\varepsilon>0$ and black-box access to an $n$-qubit unitary $U$.}
\medskip

\For{$m=O(1/\varepsilon)$ independent trials}{
    Sample a uniformly random $n$-qubit stabilizer state $\ket{S}$\;
    Run the tester from \cref{thm:guarantees-stab-tester} 
    on copies of $U\ket{S}$ with error parameter $\delta=O(\eps)$\;
}
\lIf{all $m$ trials accept}{\Return{\textbf{Accept}}}
\lElse{\Return{\textbf{Reject}}}
\end{algorithm}
\medskip

If $U$ is a Clifford unitary, then $U\ket{S}$ is a stabilizer state for every stabilizer input state $\ket{S}\in \mathrm{Stab}(n)$.  Hence, our Clifford testing algorithm directly inherits the completeness guarantees above. Soundness requires additional work: we must relate the Clifford fidelity of $U$ to the expected stabilizer fidelity of $U\ket{S}$ over random $\ket{S}$. In other words, we need to show that if $U$ is far from any Clifford (low Clifford fidelity), then with high probability over $\ket{S}$, the output $U\ket{S}$ is far from every stabilizer state. Establishing this relationship is the main technical contribution of this section. To this end, below we prove the following theorem:
\begin{thm}
[Bounds on average stabilizer fidelity of $U\ket{S}$]\label{thm:average-stabilizer-fidelity}
Let $U$ be an $n$-qubit unitary and let $\ket{S}$ be a uniformly random $n$-qubit stabilizer state. Then, it holds that
\begin{equation}
F_{\mathrm{Cliff}}(U)\leq \underset{\ket S\in\mathrm{Stab}\left(n\right)}{\mathbb{E}}\big [F_{\mathrm{Stab}}\left(U\ket S\right)\big]\leq\bigg(\frac{1}{8}F_{\mathrm{Stab}}(\kett{U}) + \frac{7}{8} +9\cdot 2^{-n} \bigg)^{1/4},
\end{equation}
\end{thm}
Recall from \cref{thm:equivalence-fcliff-fstab} that $F_{\mathrm{Stab}}(\kett{U}) =F_{\mathrm{Cliff}}(U) $ whenever $F_{\mathrm{Cliff}}(U) >1/2$.
In particular, the upper bound ensures that Clifford fidelity bounded away from $1$ implies detectably low average stabilizer fidelity, up to an exponentially small correction in $n$.
As a corollary, we obtain an auxiliary-free, non-adaptive, single-copy Clifford tester.

\auxfreesingle*

\begin{proof} 
Denote by $m$ the number of independent trials (we will fix its value later). In each trial, a new independent $\ket{S}\in \mathrm{Stab}(n)$ is drawn, and we run the single-copy stabilizer tester using $t_{\text{per trial}}=O \big(\frac{n}{\varepsilon^2} \log \frac{1}{\delta} \big)$ copies of $U\ket{S}$, where we choose $\delta = 1/(3m)$.

To argue about completeness, assume 
$U$ is a Clifford unitary. Then, for any $\ket{S}\in \mathrm{Stab}(n)$, $U\ket{S}$ is a stabilizer state. Then, by \cref{thm:guarantees-stab-tester}, we have that per trial, the failure probability is $\delta$ and so by a union bound over the $m$ independent trials, we find 
that
\begin{equation}
     \Pr[ \mathrm{accept}] \geq 1- m/\delta = 2/3.
\end{equation}
To argue about soundness, assume $F_{\mathrm{Cliff}}(U)\leq 1-\varepsilon$ with $\varepsilon<1/2$, then by \cref{thm:average-stabilizer-fidelity}, we have
\begin{equation}
    \underset{\ket S\in\mathrm{Stab}\left(n\right)}{\mathbb{E}}\big [F_{\mathrm{Stab}}\left(U\ket S\right)\big]\leq\ 1-\Omega(\varepsilon)+O(2^{-n})  .
\end{equation}
By Markov's inequality, for sufficiently large $n$, with probability $p:=\Omega(\varepsilon)$ over the random choice of $\ket{S}$, we have $F_{\mathrm{Stab}}\left(U\ket S\right)\leq\ 1-\Omega(\varepsilon)$. By independence, a single trial hence detects the non-Cliffordness with probability $p\;(1-\delta)$.
The probability that at least a single out of the $m$ trials detects non-Cliffordness is 
\begin{equation}
    \Pr[ \mathrm{reject}] \geq 1 - \exp(-m \;p(1-\delta)) = 1- \exp(-pm + p/3).
\end{equation}
 Hence, choosing $m$ such that
\begin{equation}
    -pm + p/3 \leq \ln (1/3) \Leftrightarrow m \geq \frac{\ln (3)}{p} + 1/3
\end{equation}
is sufficient to guarantee soundness. Since, $p=\Omega(\varepsilon)$, we find that asymptotically the choice $m = O\big(\frac{1}{\varepsilon} \big)$ is sufficient.
The total query complexity is hence
\begin{equation}
    m \cdot t_{\text{per trial}} = O \bigg(\frac{1}{\varepsilon}\bigg)  \cdot O \bigg(\frac{n}{\varepsilon^2} \log \frac{1}{\varepsilon}   \bigg ) = O \bigg(\frac{n}{\varepsilon^3} \log \frac{1}{\varepsilon} \bigg).
\end{equation}
Similarly, the total time complexity is $ m \cdot \mathrm{time}_{\text{per trial}} = O(1/\varepsilon) \cdot O\big(\frac{n^3}{\varepsilon^2} \log \frac{1}{\epsilon}\big)$.

\end{proof}

In the remainder of this section, we will prove \cref{thm:average-stabilizer-fidelity}.

\begin{proof}[Proof of \cref{thm:average-stabilizer-fidelity}]
For any $n$-qubit state $\ket{\psi}$, by \cref{cor:upperbound-fidelity},
\begin{equation}\label{eq:upperbound}
F_{\mathrm{Stab}}\left(\ket{\psi}\right)\leq\left(2^n\sum_{x}p_{\ket{\psi}}\left(x\right)^{2}\right)^{1/4}=\left(2^n\left\Vert p_{\ket{\psi}}\right\Vert _{2}^{2}\right)^{1/4}.
\end{equation}
Hence, it also holds on average over all stabilizer states:
\begin{equation}
\underset{\ket S\in\mathrm{Stab}\left(n\right)}{\mathbb{E}}F_{\mathrm{Stab}}\left(U\ket S\right)\leq\underset{\ket S\in\mathrm{Stab}\left(n\right)}{\mathbb{E}}\left(2^n\left\Vert p_{U\ket S}\right\Vert _{2}^{2}\right)^{1/4}.
\end{equation}
Next, since $f(x)=x^{1/4}$ is concave, we can use Jensen's inequality
to get
\begin{equation}
\underset{\ket S\in\mathrm{Stab}\left(n\right)}{\mathbb{E}}F_{\mathrm{Stab}}\left(U\ket S\right)\leq\left(2^n\underset{\ket S\in\mathrm{Stab}\left(n\right)}{\mathbb{E}}\left\Vert p_{U\ket S}\right\Vert _{2}^{2}\right)^{1/4}.
\end{equation}
Now writing out the 2-norm, we have
\begin{equation}
\underset{\ket S\in\mathrm{Stab}\left(n\right)}{\mathbb{E}}\left[\left\Vert p_{U\ket S}\right\Vert _{2}^{2}\right]=\underset{\ket S\in\mathrm{Stab}\left(n\right)}{\mathbb{E}}\trace\left(U^{\dagger\otimes4}\Pi_{4}U^{\otimes4}\,\ket S\bra S^{\otimes4}\right).
\end{equation}
Next, we can use \cref{fact:four-copy-average-stabilizer-state} which we restate here for convenience
as
\begin{align}
\mathbb{E}_{\ket S}\ket S\bra S^{\otimes4} & =\frac{1}{2^n \,D_{+}}\left(\Pi_{4}\Pi_{\mathrm{sym}}+\frac{4}{\left(2^n +4\right)}\left(I-\Pi_{4}\right)\Pi_{\mathrm{sym}}\right)\\
\nonumber
 & =\frac{1}{2^n \,D_{+}}\left(\frac{4}{\left(2^n +4\right)}\Pi_{\mathrm{sym}}+\left(1-\frac{4}{\left(2^n +4\right)}\right)\Pi_{4}\Pi_{\mathrm{sym}}\right),
\end{align}
where $D_{+}=\frac{\left(2^n +1\right)\left(2^n +2\right)}{6}=\trace\left(\Pi_{4}\Pi_{\mathrm{sym}}\right)$. With this fact we can calculate 
\begin{align}
\underset{\ket S\in\mathrm{Stab}\left(n\right)}{\mathbb{E}}\left[\left\Vert p_{U\ket S}\right\Vert _{2}^{2}\right] & =\frac{1}{2^n \,D_{+}}\left(\frac{4}{\left(2^n +4\right)}\trace\left(\Pi_{4}\Pi_{\mathrm{sym}}\right)+\left(1-\frac{4}{\left(2^n +4\right)}\right)\trace\left(U^{\dagger\otimes4}\Pi_{4}U^{\otimes4}\Pi_{4}\Pi_{\mathrm{sym}}\right)\right)\notag\\
 & =\frac{1}{2^n\left(2^n +4\right)}\left(4+\frac{2^n }{D_{+}}\trace\left(U^{\dagger\otimes4}\Pi_{4}U^{\otimes4}\Pi_{4}\Pi_{\mathrm{sym}}\right)\right).\label{eq:proof_1}
\end{align}
Let us focus on the trace term in the final equation. We expand the projector onto the symmetric subspace into permutations, $\Pi_{\mathrm{sym}}=\frac{1}{4!}\sum_{\pi \in \mathcal{S}_4}R(\pi)$, to get
\begin{equation}
\trace\left(U^{\dagger\otimes4}\Pi_{4}U^{\otimes4}\Pi_{4}\Pi_{\mathrm{sym}}\right) = \frac{2^{-4n}}{24} \sum_{\pi\in S_4} \sum_{x,y \in \mathbb{F}_2^{2n}} \trace\big((U\ct P_xU P_y)\tn{4} R(\pi)\big).
\end{equation}
Each term in the sum over permutations depends only on the cycle type of the permutation. We begin by evaluating the cycle type $(1,1,1,1)$ corresponding to the identity permutation $\pi=e$ with $R(e)=I^{\otimes 4}$,
\begin{equation}
2^{-4n}\sum_{x,y \in \mathbb{F}_2^{2n}} \trace\big(U\ct P_xU P_y\big)^4 = 2^{2n}\cdot \big( 2^{2n} \lVert p_U \rVert_2^2 \big)  \leq 2^{2n}\frac{1+ F_{\mathrm{Stab}}(\kett{U})}{2}.
\end{equation}
This last inequality is due to \cref{lem:bound-on-acceptance-probability}.
Next we evaluate the $(2,2)$ cycle type, which has $3$ elements:
\begin{equation}
2^{-4n}\sum_{x,y \in \mathbb{F}_2^{2n}} \trace\big((U\ct P_xU P_y)^2\big)^2 \leq 2^{-4n}\sum_{x,y \in \mathbb{F}_2^{2n}}2^{2n} = 2^{2n}.
\end{equation}
It will turn out that the contributions due to all the other cycle types are sub-leading. We have for the $(3,1)$ cycle type:
\begin{align}
    2^{-4n}\sum_{x,y \in \mathbb{F}_2^{2n}} \trace\big((U\ct P_xU P_y)^3\big)&\trace\big(U\ct P_xU P_y\big)\\&\leq 2^{-4n} \max_{x,y \in \mathbb{F}_2^{2n}} |\trace\big((U\ct P_xU P_y)^3\big)| \sum_{x,y \in \mathbb{F}_2^{2n}} |\trace\big(U\ct P_xU P_y\big)|
     \nonumber\\
    \nonumber
    & \leq 2^{-4n} \cdot 2^n \cdot 2^{4n} = 2^n,
     \nonumber
\end{align}
where we have used a 1-norm to 2-norm bound $\lVert \cdot \rVert_1 \leq \sqrt{d} \lVert \cdot \rVert_2$ with $d$ being the dimension of the vector space such that
\begin{equation}
    \sum_{x,y \in \mathbb{F}_2^{2n}} |\trace\big(U\ct P_xU P_y\big)| \leq \sqrt{2^{4n}} \sqrt{\sum_{x,y \in \mathbb{F}_2^{2n}} |\trace\big(U\ct P_xU P_y\big)|^2} = \sqrt{2^{4n}} \sqrt{2^{4n}} = 2^{4n}.
\end{equation}

\noindent For the $(4)$ cycle type: 
\begin{equation}
2^{-4n}\sum_{x,y \in \mathbb{F}_2^{2n}} \trace\big((U\ct P_xU P_y)^4\big) \leq 2^{-4n}\sum_{x,y \in \mathbb{F}_2^{2n}}2^{n}=2^{n},
\end{equation}
and finally for the $(2,1,1)$ cycle type,
\begin{align}
2^{-4n}\sum_{x,y \in \mathbb{F}_2^{2n}} \trace\big((U\ct P_xU P_y)^2\big)&\trace\big(U\ct P_xU P_y\big)^2\notag\\
&\leq 2^{-4n} \max_{x,y \in \mathbb{F}_2^{2n}} |\trace\big((U\ct P_xU P_y)^2\big)|\sum_{x,y \in \mathbb{F}_2^{2n}} \trace\big(U\ct P_xU P_y\big)^2\\
&\leq 2^{n}.
 \nonumber
\end{align}
This means that the contribution of all permutations in $\mathcal{S}_4$ with cycle type different from $(1,1,1,1)$ or $(2,2)$ (of which there are $20$) can be jointly upper bounded by $2\cdot2^{n}$. Combining all contributions, we get
\begin{equation}
\trace\left(U^{\dagger\otimes4}\Pi_{4}U^{\otimes4}\Pi_{4}\Pi_{\mathrm{sym}}\right) \leq \frac{2^{2n}}{24}\bigg(\frac{1+ F_{\mathrm{Stab}}(\kett{U})}{2} + 3 + 20 \cdot 2^{-n}\bigg).
\end{equation}
With this, we can finish our overall calculation, plugging in $D_{+}=\frac{\left(2^n +1\right)\left(2^n +2\right)}{6}$, to obtain 
\begin{align}
\underset{\ket S\in\mathrm{Stab}\left(n\right)} 
{\mathbb{E}}F_{\mathrm{Stab}}\left(U\ket S\right)
&\leq\left(2^n\underset{\ket S\in\mathrm{Stab}\left(n\right)}{\mathbb{E}}\left\Vert p_{U\ket S}\right\Vert _{2}^{2}\right)^{1/4}\\
 \nonumber
 &= \left(\frac{2^n }{2^n\left(2^n +4\right)}\left(4+\frac{2^n }{D_{+}}\trace\left(U^{\dagger\otimes4}\Pi_{4}U^{\otimes4}\Pi_{4}\Pi_{\mathrm{sym}}\right)\right)  \right)^{1/4}\\
  \nonumber
&\leq \left(\frac{1}{\left(2^n\!+\!4\right)}\left(4+\frac{2^{3n}}{(2^n\!+\!1)(2^n\!+\!2)}\bigg(\frac{1}{8}F_{\mathrm{Stab}}(\kett{U}) + \frac{7}{8} + 5\cdot 2^{-n}\bigg)\right)\right)^{1/4}\\
 \nonumber
&\leq \left(\frac{1}{2^{n}}\left(4+\frac{2^{3n}}{2^{2n}}\bigg(\frac{1}{8}F_{\mathrm{Stab}}(\kett{U}) + \frac{7}{8} + 5\cdot 2^{-n}\bigg)\right)\right)^{1/4}\\
 \nonumber
&=\bigg(\frac{1}{8}F_{\mathrm{Stab}}(\kett{U}) + \frac{7}{8} + 9\cdot 2^{-n} \bigg)^{1/4},
 \nonumber
\end{align}
which proves the upper bound in the theorem statement. 
It remains to prove the associated lower bound. 
We have from the definition of $F_{\mathrm{Cliff}}(U)$:
\begin{align}
    F_{\mathrm{Cliff}}(U) = 2^{-2n} \max_{C\in \mathrm{Cliff}} |\trace(U\ct C)|^2 &= \max_{C\in \mathrm{Cliff}}\big|\mathbb{E}_{\ket{S}} \bra{S}U\ct C\ket{S}\big|^2\\
    &\leq \big|\mathbb{E}_{\ket{S}}\max_{S'}\bra{S}U\ct\ket{S'}\big|^2 \leq \mathbb{E}_{\ket{S}}F_{\mathrm{Stab}}(U\ket{S}).
     \nonumber
\end{align}
This completes the proof.
\end{proof}

\section{Single-copy lower bounds}
\label{sec:single-copy-lower-bounds}
This section is organized as follows: In \cref{ssec:tree-rep-framework}, we review the tree representation framework \cite{bubeckEntanglementNecessaryOptimal2020,aharonovQuantumAlgorithmicMeasurement2022,chenExponentialSeparationsLearning2022a} for modelling adaptive single-copy algorithms in the context of channel discrimination tasks.

In \cref{ssec:reduction-to-depol}, we argue that a query lower bound for single-copy Clifford testing can be obtained by considering the task of distinguishing a uniformly random Clifford unitary channel from the completely depolarizing channel. This task is somewhat analogous to that of distinguishing a Haar random unitary channel from the completely depolarizing channel, considered in Refs.\ \cite{aharonovQuantumAlgorithmicMeasurement2022,chenExponentialSeparationsLearning2022a}. The main difference is that the unitary group is replaced with the Clifford group. The reduction also parallels the one from Ref.~\cite{hinscheSingleCopyStabilizerTesting2025} for single-copy stabilizer testing.

In \cref{ssec:partial-transposes}, we present our novel results on the structure of the
partial transposes of the Clifford commutant generators.

These results are then used in \cref{ssec:auxiliary-free-lower-bound}, where we establish the main result of this section, a $\Omega(n^{1/4})$ query complexity lower bound for single-copy Clifford testing. This bound holds against auxiliary-free testers, i.e.,  those that do not have access to an auxiliary register.

Finally, in \cref{ssec:auxiliary-assisted-lower-bound}, we explain why this proof strategy based on partial transposes does not extend to the auxiliary-assisted setting.

\subsection{Tree representation framework}\label{ssec:tree-rep-framework}
To prove our lower bounds, we will be interested in distinguishing tasks of the following form.

\begin{definition}[$t$-query channel distinguishing task]
Let $\mu$ and $\nu$ be two ensembles of quantum channels, i.e., CPTP maps $\mathcal{E}:\mathcal{L}(\mathcal{H}_{\mathrm{main}})\to\mathcal{L}(\mathcal{H}_{\mathrm{main}})$. We consider the following two events to
happen with equal probability of 1/2: 
    \begin{itemize}
        \item The unknown channel $\mathcal{E}$ is sampled according to $\mu$.
        \item The unknown state $\mathcal{E}$ is sampled according to $\nu$.
    \end{itemize}
  Given access to $t$ queries of the unknown channel $\mathcal{E}$, the goal is to design a quantum algorithm (i.e.,  some physical quantum experiment) that decides correctly between these two events with probability $\geq 2/3$.
\end{definition}

Throughout, we fix the number of queries to be $t$. We will be interested in $n$-qubit channels so that $\mathcal{H}_{\mathrm{main}}=\mathbb{C}^{2^n}$.

Following the framework established in Ref.\ \cite{chenExponentialSeparationsLearning2022a}, we model (possibly adaptive) single-copy channel testing protocols using the \textit{tree representation framework}. Therein, a single-copy algorithm for a $t$-query distinguishing task is represented by a rooted tree $\mathcal{T}$ of depth $t$ where every node corresponds to a prepare-and-measure experiment. That is, the algorithm prepares an input quantum state (possibly entangled with an auxiliary system $\mathcal{H}_{\mathrm{aux}}$), passes it through the channel and makes a POVM measurement on the output. After the experiment, the state of the algorithm moves to a child
node of $u$ depending on the experimental outcome $s$ obtained so that each node corresponds to a transcript of prior measurement outcomes. This tree structure naturally accommodates adaptive protocols where input states and measurements can depend on previously obtained measurement outcomes.

To formalize this, let us set up some notation. We identify each node of the tree $\mathcal{T}$ with its transcript of outcomes, i.e., $u_i = (s_1,\dots,s_i)$ for $0\leq i \leq t$.
\begin{enumerate}
    \item The root node is denoted $u_0=\varnothing$.
    \item At each node $u$, the protocol specifies
    \begin{itemize}
        \item an input state $\rho^u \in \mathcal{L}(\mathcal{H}_{\mathrm{main}}\otimes \mathcal{H}_{\mathrm{aux}})$, and
        \item a POVM $\{M^u_s\}_{s}$ acting on $\mathcal{H}_{\mathrm{main}}\otimes \mathcal{H}_{\mathrm{aux}}$.
    \end{itemize}
    \item Let $\mathcal{E}:\mathcal{L}(\mathcal{H}_{\mathrm{main}})\to\mathcal{L}(\mathcal{H}_{\mathrm{main}})$ denote the unknown channel and let $\mathcal{I}_{\mathrm{aux}}$ be the identity channel on the auxiliary space $\mathcal{H}_{\mathrm{aux}}$.  
    The conditional probability of observing outcome $s_i$ in round $i$, given the previous transcript $u_{i-1}=(s_1,\dots,s_{i-1})$, is
    \begin{equation}
        \Pr(s_i \mid u_{i-1}) = 
        \tr\Big[\,M^{u_{i-1}}_{s_i}\,
        \big(\mathcal{E}\otimes \mathcal{I}_{\mathrm{aux}}\big)(\rho^{u_{i-1}})\,\Big].
    \end{equation}
   We note that each $\{M_{s_i}^{u_{i-1}}\}_{s_i}$ forms a POVM since $\sum_{s_i} M_{s_i}^{u_{i-1}} = I$.  
    \item The leaves of the tree $\mathcal{T}$ correspond to complete transcripts across all $t$ rounds, i.e., $\ell=(s_1,\dots,s_t)$.  
    By the chain rule, the probability of reaching a 
    leaf $\ell$ under channel $\mathcal{E}$ is
    \begin{align}
        p_{\mathcal{E}}(\ell)
        &= \Pr(s_1)\Pr(s_2 | u_1) \cdots \Pr(s_t |u_{t-1}) \\
        &=
        \prod_{i=1}^t 
        \tr\Big[\,M^{u_{i-1}}_{s_i}\,
        \big(\mathcal{E}\otimes \mathcal{I}_{\mathrm{aux}}\big)(\rho^{u_{i-1}})\,\Big].\nonumber
    \end{align}
We will use the notation 
    \begin{align}
    \rho_{\ell} &:=\bigotimes_{i=1}^t\rho^{u_{i-1}} \in \mathcal{L}(\mathcal{H}_{\mathrm{main}}\otimes \mathcal{H}_{\mathrm{aux}})^{\otimes t}, \\
    M_{\ell}&:= \otimes_{i=1}^t M^{u_{i-1}}_{s_i} \in \mathcal{L}(\mathcal{H}_{\mathrm{main}}\otimes \mathcal{H}_{\mathrm{aux}})^{\otimes t}
\end{align}
for the states and POVM elements along a root-to-leaf path.
Then, we can rewrite the leaf probabilities simply as
\begin{equation}
     p_{\mathcal{E}}(\ell) = \tr\Big[\,M_{\ell} \big(\mathcal{E}\otimes \mathcal{I}_{\mathrm{aux}}\big)^{\otimes t} \big(\rho_{\ell}\big) \,\Big].
\end{equation}
    By writing out the sum over leaves as a nested sum, 
   \begin{equation}
     \sum_{\ell \in \mathrm{leaf}(\mathcal{T})}M_{\ell}  = \sum_{s_1} \cdots \sum _{s_{t-1}}\sum _{s_t}\;\bigotimes_{i=1}^t M_{s_i}^{u_{i-1}} = \sum_{s_1}\sum _{s_{t-1}} \bigotimes_{i=1}^{t-1} M_{s_i}^{u_{i-1}}\otimes \underbrace{\sum  _{s_t} M_{s_t}^{u_{t-1}}}_{=I} = \cdots=I^{\otimes t},
   \end{equation}
   we see that $\{M_{\ell}\}_{\ell}$ forms a POVM on the $t$ copies of $\mathcal{H}_{\mathrm{main}}\otimes \mathcal{H}_{\mathrm{aux}}$.
\end{enumerate}

This summarizes the notation we will use in the context of the tree representation framework. To show single-copy query-complexity lower bounds in this framework, the starting point is Le Cam's two-point method (see, e.g., Ref.\ \cite[Section 3.1]{canonneTopicsTechniquesDistribution2022}).

\begin{lem}[Le Cam's two-point method]
    The probability that the
distinguishing algorithm corresponding to a tree $\mathcal{T}$ solves the two-hypothesis channel distinction task correctly is upper bounded by the total variation distance of the distributions over the leaves,
\begin{equation}
   \left\lVert \underset{ \mathcal{E}\sim \mu}{\mathbb{E}} [p_{\mathcal{E}}]- \underset{ \mathcal{E}\sim \nu}{\mathbb{E}} [p_{\mathcal{E}}]  \right\rVert_{\mathrm{TV}}= \frac{1}{2} \sum_{\ell \in \mathrm{leaf}(\mathcal{T})} \left| \underset{ \mathcal{E}\sim \mu}{\mathbb{E}} [p_{\mathcal{E}}(\ell)] - \underset{ \mathcal{E}\sim \nu}{\mathbb{E}} [p_{\mathcal{E}}(\ell)] \right|.
\end{equation}
\end{lem}

\subsection{Reduction to distinguishing a random Clifford from the completely depolarizing channel}\label{ssec:reduction-to-depol}
First, we argue that we can prove single-copy lower bounds for Clifford testing by proving single-copy lower bounds for a particular distinguishing task, namely that of distinguishing a random Clifford unitary from the completely depolarizing channel.
This reduction is essentially analogous to the one used in 
Ref.\ \cite{hinscheSingleCopyStabilizerTesting2025} to prove single-copy lower bounds for stabilizer testing via the distinguishing task of a random stabilizer state vs the maximally mixed state.

To establish this reduction, we consider the following three ensembles of channels that the algorithm has access to:
\begin{enumerate}[label=\textbf{(\alph*)}]
    \item[(H)] Haar random $n$-qubit unitaries, $\mathcal{U}(\cdot)=U(\cdot)U^{\dagger}$ with $U\sim \mu_H$ denoting the Haar measure on the $n$-qubit unitary group $\mathrm{U}(2^n)$,
    \item[(C)] uniformly random $n$-qubit Clifford unitaries, $\mathcal{C}(\cdot)=C(\cdot)C^{\dagger}$ with $C\sim \Cln$ sampled uniformly at random from the $n$-qubit Clifford group $\Cln$,
    \item[(D)] the completely depolarizing $n$-qubit channel,  $\mathcal{D} := \tr(\cdot) I/2^{n}$.
\end{enumerate}
For each pair of ensembles, we can consider an associated distinguishing task. For instance, the pair $(H, C)$ corresponds to distinguishing a Haar random unitary from a uniformly random Clifford unitary. This task is also the natural starting point for proving a lower bound on the sample complexity of Clifford testing since it can be reduced to Clifford testing: With overwhelming probability, a Haar random unitary $U$ is far from all Clifford unitaries with respect to our distance measure, the Clifford in-fidelity $1-F_{\mathrm{Cliff}}(U)$ (c.f. \cref{eq:clifford-fidelity}). Hence, any Clifford testing algorithm would likely reject it but accept a uniformly random Clifford unitary. This observation is formalized via the following lemma:
\begin{lem}\label{lem:reduction-to-haar-random-vs-clifford}
    Let $0<\varepsilon < 1-\Omega(n^2/2^{2n})$. Then, any algorithm for Clifford testing to accuracy $\varepsilon$ using $t$ queries can solve the $t$-query distinguishing task of deciding between a Haar random $n$-qubit unitary and a uniformly random $n$-qubit Clifford unitary with probability $1-2^{-O(n^2)}$.
\end{lem}

This reduction, i.e., applying a Clifford testing algorithm to distinguish between the Haar random ensemble and uniformly random Clifford ensemble, may fail with a small probability, as indicated in \Cref{lem:reduction-to-haar-random-vs-clifford}, namely when the Haar randomly sampled unitary happens to be $\varepsilon$-close to a Clifford unitary. In this event, it is not guaranteed that a Clifford testing algorithm correctly distinguishes the two ensembles. To prove \Cref{lem:reduction-to-haar-random-vs-clifford}, we hence bound the probability of this event as follows:
\begin{fact}[Probability that Haar random unitary is $\varepsilon$-close to Clifford] \label{fact:haar-close-to-Clifford}
Let $0<\varepsilon < 1- \Omega( n^2/2^{2n})$. Then, for a Haar random $n$-qubit unitary $U$,
    \begin{equation}
        \Pr_{U\sim \mu_H}\left[\max_{C\in \Cln}\left|\braket{\!\langle C}{U\rangle\!}\right|^{2}\geq1-\varepsilon\right]\leq 2^{-O(n^2)}\,.
    \end{equation}
\end{fact}
\begin{proof}
    This bound is a consequence of Levy's lemma (see,  e.g.,  
    Ref.\ \cite{ledouxConcentrationMeasurePhenomenon2001})
    leading to exponential concentration, combined with a union bound over all Clifford unitaries. Concretely, using the results of Ref.~ \cite[Lemma 3.2]{lowLargeDeviationBounds2009}, we have that for an $L$-Lipschitz function on $\mathrm{U}(d)$, it holds that
     \begin{equation}\label{eq:concentration-levy}
         \Pr_{U\sim \mu_H}\left[|f(U)- \mathbb{E}f(U)|\geq \delta\right] \leq 4\exp \left(-C_1\frac{d \,\delta^2}{L^2} \right), \quad \text{with } C_1 = 2/(9\pi^3).
     \end{equation}
    Here, Lipschitz continuity is measured with respect to the Hilbert-Schmidt norm.
    To use this, let $\ket{\phi}\in\mathbb{C}^{2^{2n}}$ be a fixed $2n$-qubit pure state and define $f:U(d)\to \mathbb{R}$ to be $f(U):=|\braket{\phi}{U\rangle\!}|$. This function has Lipschitz constant $L\leq 1/d^{1/2}$ with respect to the Hilbert-Schmidt norm and its mean is bounded as $\mathbb{E}[f(U)]\leq 1/d$. Then, using \cref{eq:concentration-levy}, the probability of a Haar random Choi state $\ket{U\rangle\!}$ being $\varepsilon$-close in fidelity to the fixed state $\ket{\phi}$ can be bounded as 
\begin{equation}
    \Pr_{U\sim\mu_H}\left[\left|\braket{\phi}{U\rangle\!}\right|^{2}\geq1-\varepsilon\right]\leq 4\exp\left(-C_1 d^2\left(\sqrt{1-\varepsilon}-\tfrac{1}{d}\right)^2\right) \leq4\exp\left(-C_1 \frac{d^2 (1-\varepsilon)}{4}\right),
\end{equation}
where the last inequality holds for $\varepsilon \leq 1- 4/2^{2n}$.
The number of Clifford unitaries is $\left|\Cln\right|=2^{O(n^2)}$. The result now follows from the union bound.
\end{proof}

Next, we argue that, when considering single-copy algorithms, any sample complexity lower bound for distinguishing a uniformly random Clifford unitary from the completely depolarizing channel (the pair $(C, D)$) leads to a lower bound for the pair $(H, C)$. 
This essentially follows from a triangle inequality between the three pairs as we now explain: Consider an arbitrary single-copy distinguishing algorithm using $t$ queries to the unknown channel $\mathcal{E}$. This algorithm may be represented by a tree $\mathcal{T}$ and associated a distribution $p_\mathcal{E}$ over leaves. Then, we 
apply the triangle inequality to the total variation distance between leaf distributions as
\begin{equation}\label{eq:triangle-inequality}
      \left\lVert \underset{ U\sim \mu_H}{\mathbb{E}} [p_{\mathcal{U}}]- \underset{ C\sim \Cln}{\mathbb{E}} [p_{\mathcal{C}}]  \right\rVert_{\mathrm{TV}} 
      \leq
     \left\lVert \underset{ \mathcal{U}\sim \mu_H}{\mathbb{E}} [p_{\mathcal{U}}]- p_{\mathcal{D}}  \right\rVert_{\mathrm{TV}} 
     +  \left\lVert p_{\mathcal{D}}- \underset{ C\sim \Cln}{\mathbb{E}} [p_{\mathcal{C}}]  \right\rVert_{\mathrm{TV}}.
\end{equation}
In Ref.\ \cite{chenExponentialSeparationsLearning2022a}, the authors proved single-copy lower bounds for the distinguishing task corresponding to the pair $(H,D)$. In particular, they proved the following lower bound on the TV distance of the leaf distributions:
\begin{thm}[Bound for Haar random unitaries vs. depolarizing, Theorem 7.9 in Ref.\ \cite{chenExponentialSeparationsLearning2022a}]\label{thm:single-copy-lower-bound-haar-random}
Consider an arbitrary auxiliary-assisted, adaptive single-copy algorithm for distinguishing Haar random unitaries $U\sim\mathrm{U}(d)$ from the completely depolarizing channel $\mathcal{D}$ on $\mathbb{C}^d$ using $t$ queries. Let $\mathcal{T}$ denote the associated tree. Then, for $t\leq \sqrt{d}$, the total variation distance of the leaf distributions of $\mathcal{T}$ is upper bounded as follows,
    \begin{equation}
        \left \lVert \underset{ \mathcal{U}\sim \mu_H}{\mathbb{E}} [p_{\mathcal{U}}]-p_{\mathcal{D}}  \right\rVert_{\mathrm{TV}} \leq O\bigg(\frac{t^3}{d}\bigg).
    \end{equation}
    Here, $\mu_H$ denotes the Haar measure on $\mathrm{U}(d)$.
\end{thm}
Hence, we can provide an upper bound to the total variation distance on the LHS of \cref{eq:triangle-inequality} by providing an upper bound to $\left\lVert p_{\mathcal{D}}- \mathbb{E}_{C\sim \Cln} [p_{\mathcal{C}}]  \right\rVert_{\mathrm{TV}}$.
This is why, throughout the rest of this entire section, we will focus on the pair $(C,D)$ which corresponds to the following task:
\begin{definition}[$t$-query Clifford distinguishing problem]\label{def:distinguishing_cliffords}
The following two events happen with equal prior
probability of $1/2$: 
\begin{enumerate}
\item The unknown channel $\mathcal{E}$ corresponds to a uniformly random $n$-qubit Clifford unitary. That is, it is of the form $\mathcal{C}(\cdot)=C (\cdot)C^{\dagger}$, where $C$ is drawn uniformly at random from $\Cln$.
\item The unknown channel $\mathcal{E}$ is the completely depolarizing channel $\mathcal{D}= \tr(\cdot) I/2^{n}$ on $n$ qubits.
\end{enumerate}
Given access to $t$ queries of the unknown channel $\mathcal{E}$, decide correctly between these two events with probability $\geq 2/3$.
\end{definition}

\begin{remark}
     This $t$-query Clifford distinguishing problem constitutes an instance of a problem in multi-use, binary quantum channel discrimination.
\end{remark}

\subsection{Partial transposes of commutant generators}\label{ssec:partial-transposes}
In this section, we study how the generators $\{R(T)\}_{T \in \Sigma_{t,t}}$ of the Clifford commutant behave under partial transpose operations. Understanding these transformations will be key for analyzing overlaps with operators that remain positive under partial transposes (PPT operators). We will leverage this in the next subsection to establish our single-copy lower bound.   

For an operator $A \in \mathcal{L}(((\mathbb{C}^{2})^{\otimes n})^{\otimes t})$ on $t$ copies of an $n$-qubit Hilbert space and a subset $S \subset [t]$, we denote by $A^{\Gamma_S}$ the partial transpose of $A$ with respect to the subsystems indexed by $S$.  

In earlier work~\cite{hinscheSingleCopyStabilizerTesting2025} by some of the authors, it was shown that every nontrivial generator $R(T)$ admits a non-unitary partial transpose:

\begin{thm}[Non-unitary partial transposes]
\label{thm:non-unitary-partial-transposes}For all $T\in\Sigma_{t,t} \setminus\left\{ e\right\} $,
there exists $S\subset\left[t\right]$ such $\left\Vert R(T)^{\Gamma_{S}}\right\Vert _{1}\leq2^{n(t-1)}$.
\end{thm}
In this work, we prove the complementary result: every generator $R(T)$ can also be transformed into a unitary operator by a suitable partial transpose.

\begin{thm}[Unitary partial transposes]
\label{thm:unitary-partial-transposes-rT}
For all $T\in \Sigma_{t,t}$, there exists $S\subset[t]$ such that $R(T)^{\Gamma_{S}}$ is unitary and so $\left\Vert R(T)^{\Gamma_{S}}\right\Vert _{\infty}=1$.
\end{thm}
The significance of these results lies in how the commutant generators interact with operators that remain positive under partial transposes, which we refer to as PPT (Positive Partial Transpose) operators:  
\begin{definition}[PPT operator]
    Let $A \in \mathcal{L}(((\mathbb{C}^{2})^{\otimes n})^{\otimes t})$ be a positive-semidefinite operator, i.e. $A\succeq0$ on the $t$-copy Hilbert space. We say that $A$ is a PPT operator if
    \begin{equation}
        A^{\Gamma_S} \succeq0, \quad \forall S\subset [t].
    \end{equation}
\end{definition}
The class of PPT operators includes, in particular, product and separable states as well as POVMs. 

To illustrate these implications, we use \cref{thm:unitary-partial-transposes-rT} to show a uniform bound on the overlap of PPT states and the generators of the commutant of the Clifford group:
\begin{cor}[Bound for $t$-copy PPT states]
\label{cor:tr_rt_rho_ppt_bound}
Let $\rho$ be a PPT state on $t$
copies, then for all $T\in\Sigma_{t,t}$, we have
\begin{equation}
\left|\trace\left(R(T)\rho\right)\right|\leq1.
\end{equation}
\end{cor}

\begin{proof}
By \cref{thm:unitary-partial-transposes-rT}, there exists $S\subset[t]$
such that $\lVert R\left(T\right)^{\Gamma_{S}}\rVert _{\infty}=1$
and so
\begin{equation}
\left|\trace\big(R(T)\rho\big)\right|
=\left|\trace\big(R\left(T\right)^{\Gamma_{S}}\rho^{\Gamma_{S}}\big)\right|
\leq\lVert R\left(T\right)^{\Gamma_{S}}\Vert _{\infty}\left\Vert \rho^{\Gamma_{S}}\right\Vert _{1}=1.
\end{equation}
In the last step we used the PPT assumption on $\rho$, which
implies $\lVert \rho^{\Gamma_{S}}\rVert _{1}=\trace\left(\rho^{\Gamma_{S}}\right)=\trace\left(\rho\right)=1$.
\end{proof}

The rest of this section will be devoted to explaining the structure of the partial transposes of $R(T)$ and the proof of \cref{thm:unitary-partial-transposes-rT}. Our key technical insight is to show that \cref{thm:unitary-partial-transposes-rT} can be connected to the theory of matroid intersection.

\paragraph{Stochastic Lagrangian subspaces as self-dual codes.}
Taking partial transposes does not, in general, preserve the set of Clifford commutant generators ${R(T)}_{T \in \Sigma{t,t}}$. However, it does preserve a larger set of operators ${R(D)}$ corresponding to self-dual binary codes $D$ of length $2t$. This perspective allows us to interpret partial transposes as automorphisms within a familiar coding-theoretic framework. In what follows, we make the relation to self-dual codes and orthogonal groups precise, before turning to the analysis of partial transposes in this language.

We first note that the set of stochastic Lagrangian subspaces is contained in the set of self-dual codes,
\begin{equation}
    \Sigma_{t,t} \subset \mathrm{SD}(2t),
\end{equation}
where $\mathrm{SD}(2t)$ denotes the set of all self-dual binary $[2t, t]$ codes. While this was observed in \cite{grossSchurWeylDuality2021}, we make the statement explicit here for clarity.

Indeed, the total-isotropy condition in \cref{def:sigma_tt} implies that every $T \in \Sigma_{t,t}$ is self-orthogonal with respect to the standard inner product on $\mathbb{F}_2^{2t}$
\begin{fact}[$T\in \Sigma_{t,t}$ are self-orthogonal, see Remark 4.2 in Ref.\ \cite{grossSchurWeylDuality2021}]
For all $T\in \Sigma_{t,t}$, we have $T \subseteq T^\perp$, where
\begin{equation}
    T^\perp = \{ (x',y') \in \mathbb{F}^{2t}_2 \; :\; (x,y) \cdot(x',y') = 0,\; \forall (x,y)\in T\}.
\end{equation}
\end{fact}
\noindent Since each $T \in \Sigma_{t,t}$ has dimension $t$, which is the maximum dimension for a self-orthogonal subspace of $\mathbb{F}_2^{2t}$, it follows that each $T$ is in fact \emph{self-dual}, i.e., $T = T^\perp$\footnote{This also implies that every $x \in T$ has even Hamming weight, so the all-ones vector $1_{2t} := (1, \dots, 1)$ automatically lies in $T^\perp=T$.}.

For any such code $D\in \mathrm{SD}(2t)$, we can define a corresponding operator on $((\mathbb{C}^{2})^{\otimes n})^{\otimes t}$ via $R(D) = r(D)^{\otimes n}$
\begin{equation}
     r(D)= \sum_{(x,y)\in D}\ket{x}\!\bra{y}\in \mathcal{L}\big((\mathbb{C}^{2})^{\otimes t}\big).
\end{equation}
This generalizes the operators $r(T)$ introduced in \cref{thm:commutant-basis-gnw}, corresponding to the case $D=T \in \Sigma_{t,t}$.

Throughout this section, we often find it convenient to choose a basis for $D$. To this end, we let
\begin{equation}
    G = [A,B] = [a_1, \cdots, a_t \;|\; b_1, \cdots, b_t] 
\end{equation}
be a $t \times 2t$ binary generator matrix for the binary self-dual $[2t, t]$ code $D$, where $A$ denotes the left $t \times t$ block of $G$ and $B$ denotes the right $t \times t$ block.
The columns of $A$ and $B$ are denoted by $a_i$ and $b_i$, for $i\in [t]$.
The rows of $G$ form a basis for the code $D$ and each codeword $(x,y)\in D$ is of the form $u G=(uA,uB)$ for some $u\in \mathbb{F}_2^t$.

Recall from \cref{ssec:clifford-commutant} that the stochastic orthogonal group 
$\mathrm{O}_t^{(1)}$ corresponds to the unitary part of the Clifford commutant generators. 
Just as $\Sigma_{t,t}$ embeds into the larger set of self-dual codes, 
$\mathrm{O}_t^{(1)}$ embeds into the orthogonal group over $\mathbb{F}_2$, 
which precisely characterizes the unitary operators among the family 
$\{r(D)\}_{D\in \mathrm{SD}(2t)}$:

\begin{definition}[Orthogonal group]
The orthogonal group over $\mathbb{F}_2$ is defined as 
\begin{equation}
    \mathrm{O}_t = \{A\in \mathrm{GL}(t,\mathbb{F}_2) \; |\; AA^T = I\}.
\end{equation}
Equivalently, it is the group of $t\times t$ binary matrices $O$ such that
\begin{equation}\label{eq:preserving-inner-product}
    O  x \cdot O x =  x \cdot  x \mod 2 \qquad \forall  x \in \Ft \,. 
\end{equation}
\end{definition}

\noindent The defining property \Cref{eq:preserving-inner-product} implies that every row of $O \in \mathrm{O}_t$ has odd Hamming weight. In contrast, the defining property of $\mathrm{O}_t^{(1)}$ ensures that every row has Hamming weight $1 \bmod 4$. Hence, $\mathrm{O}_t^{(1)} \subset \mathrm{O}_t$. Moreover, each $O \in \mathrm{O}_t$ defines a self-dual code $D_O = \{(O x, x) \,|\, x \in \mathbb{F}_2^t\}$. Thus, $\mathrm{O}_t$ may be viewed as a subset of $\mathrm{SD}(2t)$.

The orthogonal group $\mathrm{O}_t$ gives rise to the unitary part of operators $r(D)$ for $D \in \mathrm{SD}(2t)$. 
In fact, we have the following:
\begin{fact}[Orthogonal matrices correspond to unitary operators]\label{fact:orthogonal-correspond-to-unitary}
    Let $D \in \mathrm{SD}(2t)$, then $r(D)$ is unitary if and only if $D = D_O$ for some $O\in \mathrm{O}_t$.
\end{fact}
\noindent Collecting the above observations, we arrive at the following inclusions which generalize the ones given in \cref{eq:inclusions}.
\begin{equation}  \label{eq:inclusions-updated}
\begin{array}{c c c c c}
  \mathcal{S}_t & \subset & \mathrm{O}_t^{(1)} & \subset & \Sigma_{t,t} \\
                &         & \!\!\!\cap               &         & \cap \\
                &         & \mathrm{O}_t \;      & \subset & \mathrm{SD}(2t).
\end{array}
\end{equation}
The first row corresponds to sets directly associated with generators of the Clifford commutant, while the second row corresponds to supersets preserved under partial transposes. The first and second columns highlight the unitary parts: $\mathcal{S}_t$ and $\mathrm{O}_t^{(1)}$ within the commutant, and more generally $\mathrm{O}_t$ within $\mathrm{SD}(2t)$. 

\paragraph{Partial transposes correspond to coordinate permutations.}
With this framework in place, we can now describe the effect of partial transposes on the operators $r(D)$. For any $S \subset [t]$, the partial transpose $r(D)^{\Gamma_S}$ is given by
\begin{equation}
    r(D)^{\Gamma_S} = r(D'), \quad D' = D P_S,
\end{equation}
where $P_S$ is the product of transpositions
\begin{equation}
    P_S = \prod_{i\in S} (i\; i+t),
\end{equation}
swapping coordinate $i$ with $i+t$ for each $i\in S$. Since permutations are isometries with respect to the standard inner product on $\mathbb{F}_2^{2t}$, $D'$ is again a self-dual code. This formalizes the earlier statement that partial transposes preserve the set of self-dual codes, mapping one code to another within $\mathrm{SD}(2t)$.

To describe this explicitly, let $G=[A|B]$ be a generator matrix for $D$. Then we can obtain a generator matrix $G'$ for $D'$ by swapping the columns $a_i$ with $b_i$ for each $i\in S$.

While coordinate permutations preserve the set of self-dual binary codes $\mathrm{SD}(2t)$, they do not necessarily preserve the subset $\Sigma_{t,t}$ associated with the Clifford commutant, since the total isotropy condition from \cref{def:sigma_tt} is not invariant under such swaps.

\begin{example}
Let $t=4$ and let $T_4\in \Sigma_{4,4}$ be the stochastic Lagrangian subspace with generator matrix given by $G=[A|B]$
\begin{equation}
G =
\left[
  \begin{array}{cccc|cccc}
    1 & 0 & 0 & 1 & 1 & 0 & 0 & 1\\
    0 & 1 & 0 & 1 & 0 & 1 & 0 & 1\\
    0 & 0 & 0 & 0 & 1 & 1 & 1 & 1\\
    1 & 1 & 1 & 1 & 0 & 0 & 0 & 0
  \end{array}
\right] \overset{\text{swap }a_1,b_1}{\mapsto} G' = \left[
  \begin{array}{cccc|cccc}
    1 & 0 & 0 & 1 & 1 & 0 & 0 & 1\\
    0 & 1 & 0 & 1 & 0 & 1 & 0 & 1\\
    1 & 0 & 0 & 0 & 0 & 1 & 1 & 1\\
    0 & 1 & 1 & 1 & 1 & 0 & 0 & 0
  \end{array}
\right].
\end{equation}
It is easy to check after that swapping the first column of $A$ and $B$, the resulting matrix $G'=[A'|B']$ no longer satisfies the isotropy condition, stating $x \cdot  x =  y \cdot  y \mod 4 $ for all $( x,  y) \in T$. In particular, the third and forth rows of $G'$ violate it.
\end{example}

Next, to prove \cref{thm:unitary-partial-transposes-rT}, we need to understand when the operator $r(D)$
associated with a self-dual code $D \in \mathrm{SD}(2t)$ is unitary. This will allow us to argue that there is a subset $S \subset [t]$ such that the partial transpose $R(T)^{\Gamma_S}$ becomes unitary.
Recall from \cref{fact:orthogonal-correspond-to-unitary} that $r(D)$ is unitary if and only if $D = D_O$ for some $O \in \mathrm{O}_t$. In terms of a generator matrix $G = [A|B]$ for $D$, this is equivalent to requiring that both blocks $A$ and $B$ are full rank. 

Thanks to self-duality, it actually suffices to require that either $A$ or $B$ is full rank, as $\mathrm{rank}(A) = \mathrm{rank}(B)$. This insight is implicit in Ref.\ \cite[Proposition 4.17.]{grossSchurWeylDuality2021} where it was proved for all $T\in \Sigma_{t,t}$. However, same argument applies more generally to $\mathrm{SD}(2t)$, since it only relies on self-duality.
\begin{lem}[Equal rank of $A$ and $B$]\label{lem:equal-rank-A-B}
     Let $D \in \mathrm{SD}(2t)$ be a self-dual binary $[2t,t]$ code and let $G=[A|B]$ be a generator matrix for $D$. Then, $\mathrm{rank}(A)=\mathrm{rank}(B)$.
\end{lem}

From \cref{lem:equal-rank-A-B}, the following fact immediately follows:
\begin{fact}[Unitarity of $r(D)$]\label{fact:unitarity-of-rD}
    Let $D \in \mathrm{SD}(2t)$ be a self-dual binary $[2t,t]$ code and let $G=[A|B]$ be a generator matrix for $D$. Then, $r(D)= \sum_{(x,y)\in D}\ket{x}\bra{y}$ is unitary if and only if $A$ (or equivalently $B$) is full rank.
\end{fact}

\noindent On the level of generator matrices, the partial transpose $^{\Gamma_S}$ acts by swapping the $i$-th column of $A$ with the $i$-th column of $B$ for all $i \in S$. Equivalently, after this permutation, the new left block $A'$ consists of exactly one column from each pair $\{a_i, b_i\}$, $i = 1,\dots,t$. Ensuring that $A'$ is full rank is therefore equivalent to selecting one column from each pair
\begin{equation}
    \{ a_i,\, b_i \}, \quad i = 1,\dots,t,
\end{equation}
so that the chosen columns are linearly independent. This viewpoint allows us to focus on linear independence rather than explicitly tracking which columns are swapped.
This reformulation naturally leads to a transversal problem in matroid theory.

\paragraph{Connection to matroid intersection.}
A \emph{matroid} is a mathematical structure that generalizes the concept of linear independence. A detailed exposition can be found in Ref.\ \cite{oxleyMatroidTheory2011}. Formally, we define it as follows:
\begin{definition}[Matroid]
    A matroid is a pair $M = (E, \mathcal{I})$, where $E$ is a finite \textit{ground set} and $\mathcal{I} \subseteq 2^E$ is a family of subsets of $E$ (called the \textit{independent sets}) satisfying the following axioms.
    \begin{enumerate}
        \item The empty set is independent, i.e.,  $\emptyset \in \mathcal{I}$
        \item Every subset of an independent set is independent, i.e., if $I\in \mathcal{I}$ and $I'\subseteq I$, then $I'\in \mathcal{I}$.
        \item If $I_1,I_2 \in \mathcal{I}$ are both independent and $|I_1|>|I_2|$, then there exists $x\in I_1\setminus I_2$ such that $I_2 \cup \{x \} \in \mathcal{I}$.
    \end{enumerate}
\end{definition}

One way to form a matroid is to start from a matrix:  
\begin{definition}[Vector matroid {$M[A]$}]
    Let $A$ be a matrix over a field $\mathbb{F}$. Then, the \textit{vector matroid} of $A$, denoted $M[A]$, is obtained by taking the columns as the ground set $E$ and the collection of independent sets $\mathcal{I}$ to be subsets of columns that are linearly independent over the corresponding field $\mathbb{F}$.     
\end{definition}

In our setting, we will focus on the vector matroid $M[G]$ corresponding to the generator matrix $G$ of the self-dual code $D$. In particular, we let $E = \{a_1, \dots, a_t, b_1, \dots, b_t\}$ corresponding to the $2t$ columns of $G$, and a subset $S \subseteq E$ is independent if the corresponding columns of $G$ are linearly independent over $\mathbb{F}_2$.

Now, we are interested in special subsets of columns where each columns is taken from the pair $\{a_i,b_i\}$. This pairing is captured by the concept of a \textit{transversal}, also called \textit{system of distinct representatives}:
\begin{definition}[Transversal]
 Let $E$ be a finite set. Given a family of subsets $\mathcal{X} = \{X_1, \dots, X_m\}$ of $E$, a \emph{transversal} $\mathcal{T}$ is a subset of $E$ containing exactly one element from each $X_i$.
\end{definition}
  
Here, we take $E= \{a_1, \dots, a_t, b_1, \dots, b_t\}$ and $\mathcal{X}=(X_1,\dots, X_n)$ where
\begin{equation}
    X_i = \{a_i, b_i\}, \quad i = 1, \dots, t,
\end{equation}
Then, we precisely seek a transversal $\mathcal{T}$ of $\mathcal{X}$ that is also independent in the matroid $M[G]$.  
Such a $\mathcal{T}$ corresponds exactly to choosing one column from each pair $\{a_i, b_i\}$ so that the chosen columns form a full-rank $t\times t$ matrix.

The existence of such an independent transversal is characterized by Rado's theorem:

\begin{thm}[Rado]\label{thm:rado}
Let $M = (E, \mathcal{I})$ be a matroid with rank function $r$, and let $\mathcal{X} = (X_1, \dots, X_n)$ be a family of subsets of $E$.  
Then $\mathcal{X}$ has an independent transversal in $M$ if and only if
\begin{equation}
r\left( \bigcup_{i \in I} X_i \right) \ge |I| \quad \text{for all } I \subseteq \{1, \dots, n\}.
\end{equation}
\end{thm}
For a vector matroid, the rank function $r$ coincides with the standard linear-algebraic notion of rank, i.e.,  the dimension of the subspace spanned by the columns.
For us, this theorem tells us that there is a choice of $t$ linearly
independent columns from the pairs $\left\{a_ i,b_i\right\}_{i=1}^{t}$ if and
only if for all $I\subseteq\left[t\right]$, $\mathrm{rank}\left(G_{I}\right)\geq\left|I\right|$
where $G_{I}$ is the submatrix made up of the columns $\left\{ a_i,b_i\right\}_{i\in I}$,
This is what we are going to show next.
\begin{lem}[Rank of submatrices $G_I$]\label{lem:rank-of-G_I}
Let $G=[A|B]$ be the generator matrix of a binary self-dual $[2t,t]$ code
$D$. Then, for $I\subset [t]$, let $G_I$ be the $t\times 2|I|$-submatrix of $G$ consisting of the columns $\{a_i,b_i \}_{i\in I}$ from $G$. Then,
\begin{equation}
    \mathrm{rank}(G_{I}) \geq\left|I\right|.
\end{equation} 
\end{lem}

\begin{proof}
Suppose $\mathrm{rank}(G_{I})<\left|I\right|$.
Then, 
consider the restriction of 
$D$ to the coordinates in $\left\{ a_i,b_i\right\} _{i\in I}$, $\Pi_{I}:C\to\mathbb{F}_{2}^{2\left|I\right|}$.
 Then, $\ker\Pi_{I}=C_{I}$
 with 
\begin{equation}
 D_{I}=\left\{ c\in D \;|\;c_{j}=0\quad \forall j\in\left\{ i,t+i\right\} \right\} 
\end{equation}
and we have, by rank-nullity, that $\dim C_{I}=t-\mathrm{rank}\left(G_{I}\right)>t-\left|I\right|$.

On the other hand, $C_{I}$ is self-orthogonal since it is a subspace
of the self-dual code $C$ and further $C_{I}$ can be regarded as
a code of length $2(t-\left|I\right|)$ by simply removing those all-zero
coordinates in $\left\{ i,t+i\right\} _{i\in I}$. The maximum dimension
of any binary self-orthogonal code of length $2(t-\left|I\right|)$
is 
\begin{equation}
\dim C_{I}\leq t-\left|I\right|
\end{equation}
which is a contradiction.
\end{proof}

We have now collected all the ingredients to prove our main result in this section.
\begin{restatable}[Unitary partial transposes]{thm}{unitarypartialtransposes}
\label{thm:unitary-partial-transposes-rD}
For all $D\in \mathrm{SD}(2t)$, there exists $S\subset[t]$ and $O\in \mathrm{O}_t$ such that $r(D)^{\Gamma_{S}}=r(O)$. Consequently, $r(D)^{\Gamma_{S}}$
is unitary and so $\left\Vert r(D)^{\Gamma_{S}}\right\Vert _{\infty}=1$.
\end{restatable}

\begin{proof}[Proof of \cref{thm:unitary-partial-transposes-rD}]
Let $G=[A|B]$ be a generator matrix for $D$ and let $M[G]$ be the vector matroid of $G$. Combining \cref{lem:rank-of-G_I} with Rado's theorem (\cref{thm:rado}), we conclude that there exists a transversal choice of $t$ linearly independent columns from $\{a_i, b_i\}_{i=1}^t$. 
Equivalently, there exists $S\subset[t]$ such that $D P_S =D_O$ for some $O\in \mathrm{O}_t$.
\end{proof}
\noindent Note that \cref{thm:unitary-partial-transposes-rD} is slightly more general then \cref{thm:unitary-partial-transposes-rT} in that it holds for all of $\mathrm{SD}(2t)$, i.e., for all self-dual binary codes and not just for those in $\Sigma_{t,t}$ associated with the Clifford commutant.

While our proof in this section leverages a powerful connection to matroid theory, it is not constructive. 
In \cref{sec:algorithmic-proof}, we present an algorithm that, for a given generator matrix $G$ corresponding to a code $D$, finds the partial transpose, i.e., the subset $S\subset[t]$, such that $r(D)^{\Gamma_S}=r(O)$ for some $O\in \mathrm{O}_t$.

\subsection{Lower bound against auxiliary-free, adaptive algorithms}\label{ssec:auxiliary-free-lower-bound}
In the auxiliary-free setting, the distinguishing algorithm does not have access to an auxiliary system.
For this auxiliary-free setting, we will prove the following:

\begin{thm}[Auxiliary-free TV distance bound]\label{thm:tv-distance-auxiliary-free}
Consider an arbitrary auxiliary-free, possibly adaptive, single-copy distinguishing algorithm represented by the tree $\mathcal{T}$. Let $t\leq n+1$. Then, the total variation distance between the associated leaf distributions of $\mathcal{T}$ is bounded as
    \begin{equation}
        \left\lVert p_{\mathcal{D}}- \underset{ C\sim \Cln}{\mathbb{E}} [p_{\mathcal{C}}]  \right\rVert_{\mathrm{TV}}\leq 2^{-n + O(t^4)}.
    \end{equation}
\end{thm}

By our previous discussion, in particular \cref{eq:triangle-inequality} and \cref{thm:single-copy-lower-bound-haar-random}, this bound immediately implies the following corollary.
\begin{cor}[TV distance between Haar random unitaries and random Cliffords]
Consider an arbitrary auxiliary-free, possibly adaptive, single-copy distinguishing algorithm represented by the tree $\mathcal{T}$. Let $t\leq n+1$. Then, the total variation distance between the associated leaf distributions of $\mathcal{T}$ is bounded as
    \begin{equation}
              \left\lVert \underset{ U\sim \mu_H}{\mathbb{E}} [p_{\mathcal{U}}]- \underset{ C\sim \Cln}{\mathbb{E}} [p_{\mathcal{C}}]  \right\rVert_{\mathrm{TV}}\leq 2^{-n + O(t^4)}.
    \end{equation}
\end{cor}
This corollary corresponds to a slightly more general statement than \cref{thm:Clifford-design} as it essentially considers a distance metric between the $t$-fold Haar twirl and Clifford twirl that takes into account adaptive algorithms.  
Via \cref{fact:haar-close-to-Clifford}, this implies our auxiliary-free single-copy lower bound for Clifford testing:

\begin{cor}[Lower bound for auxiliary-free, single-copy Clifford testing]
    Any auxiliary-free, possibly adaptive single-copy algorithm for Clifford testing to accuracy $0<\varepsilon< 1-\Omega(n^2/2^{2n})$ requires at least $t=\Omega(n^{1/4})$ queries.
\end{cor}

\begin{proof}[Proof of \cref{thm:tv-distance-auxiliary-free}]
To prove this, we will first write out the distribution over the leaves of the tree $\mathcal{T}$ associated with an arbitrary distinguishing algorithm.

For the random Clifford channel, we can write out the average over the Clifford group in terms of the generators of the commutant (c.f. \cref{eq:commutant_expansion}),
\begin{align}
  \underset{C\sim \Cln}{\mathbb{E}}  p_{\mathcal{C}}(\ell) 
  &=  \underset{C\sim \Cln}{\mathbb{E}}
        \prod_{i=1}^t 
        \tr
        \Big[\, M^{u_{i-1}}_{s_i}\;
        C \rho^{u_{i-1}}C^{\dagger}  \,\Big] \\
        \nonumber
   &=\sum_{T,T'\in \Sigma_{t,t}} W_{T, T'}\;   
        \tr \big[ R(T')^{\dagger} \rho_{\ell} \big]\; \tr \big[R(T) M_{\ell}  \big] \\
         \nonumber
  &=  \sum_{T,T'\in \Sigma_{t,t}} W_{T, T'}\; \vcenter{\hbox{\includegraphics[scale=1.0]{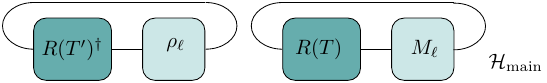}}}, \\
   \nonumber
  &:= \sum_{T,T'\in \Sigma_{t,t}} p_{T, T'} (\ell).
   \nonumber
\end{align}
On the other hand, for the completely depolarizing channel, we find
\begin{equation}
     p_{\mathcal{D}}(\ell) = \frac{1}{2^{nt}} \; \tr \big[\rho_{\ell} \big] \tr \big[ M_{\ell} \big] = \frac{\tr \big[ M_{\ell} \big]}{2^{nt}}  .
\end{equation}
By the triangle inequality, the total variation distance between the distributions over the leaves can be bounded as 
\begin{align}\label{eq:tv-distance-expansion}
   & \left\lVert p_{\mathcal{D}}- \underset{ C\sim \Cln}{\mathbb{E}} [p_{\mathcal{C}}]  \right\rVert_{\mathrm{TV}}\\
    \nonumber
    =& \;\frac{1}{2} \sum_{\ell \in \mathrm{leaf}(\mathcal{T})} \left| p_{\mathcal{D}}(\ell) -\underset{ C\sim \Cln}{\mathbb{E}} [p_{\mathcal{C}}(\ell)]  \right|  = \frac{1}{2} \sum_{\ell \in \mathrm{leaf}(\mathcal{T})} \left| p_{\mathcal{D}}(\ell) - \sum_{T,T'\in \Sigma_{t,t}} p_{T, T'}(\ell)\right | \\
     \nonumber
    \leq& \; \frac{1}{2} \sum_{\ell \in \mathrm{leaf}(\mathcal{T})} \big| p_{\mathcal{D}}(\ell) - p_{e,e}(\ell)  \big| 
    +\frac{1}{2} \sum_{\ell \in \mathrm{leaf}(\mathcal{T})} \sum_{
 T'\neq e
} |p_{e,T'}(\ell)| 
    + \frac{1}{2} \sum_{\ell \in \mathrm{leaf}(\mathcal{T})} \sum_{T\neq e, T'} |p_{T,T'}(\ell)|,
     \nonumber
\end{align}
where in the third line we have split the sum $\sum_{T,T'\in \Sigma_{t,t}}$ in a way that will turn out convenient.
In the following, we will bound each of these three terms separately. To do this, we will use asymptotic bounds on the Weingarten coefficients $|W_{T, T'}|$ stated in \cref{fact:Weingarten asymptotics}. Also note that $\big|\sum_{t,t}\big|=2^{O(t^{2})}$, so, 
e.g.,  the sum $\sum_{T\neq e,T'\in \Sigma_{t,t}}$ in the last term ranges over $2^{O(t^{4})}$ terms.

\paragraph{First term: the $(e,e)$ contribution.}
We have that
\begin{equation}
    p_{e,e}(\ell) = W_{e,e}\; \tr \big[\rho_{\ell} \big] \tr \big[ M_{\ell} \big]  = W_{e,e}\;\tr \big[ M_{\ell} \big].
\end{equation}
We can see that $(e,e)$-contribution approximately cancels with $ p_{\mathcal{D}}(\ell)$ coming from the completely depolarizing channel, since
\begin{align}
    \frac{1}{2} \sum_{\ell \in \mathrm{leaf}(\mathcal{T})} \big| p_{\mathcal{D}}(\ell) - p_{e,e}(\ell)  \big| 
    &\leq  \frac{1}{2} \bigg| \frac{1}{2^{nt}} - W_{e,e} \bigg| \underbrace{\sum_{\ell \in \mathrm{leaf}(\mathcal{T})} 
        \big| \tr \big[ M_{\ell} \big] \big|}_{=2^{nt}}\\
    & \leq \frac{1}{2} \bigg| \frac{1}{2^{nt}} - W_{e,e} \bigg| 2^{nt} \leq 2^{-n(t+1)+O(t^{2})} \cdot 2^{nt}\leq2^{-n+O(t^{2})}.
    \nonumber
\end{align}
Here, we have used that $M_l$ is positive semi-definite so that $|\tr [M_l]| = \tr [M_l]$ which lets us carry out the summation over leaves, using $ \sum_{\ell \in \mathrm{leaf}(\mathcal{T})}M_{\ell}  = I^{\otimes t}$. 

\paragraph{Second term: the $(e,T)$ contribution.}
The second term can be made uniformly small. We see by direct calculation
that
\begin{align}
     \frac{1}{2} \sum_{\ell \in \mathrm{leaf}(\mathcal{T})} \sum_{
 T' \neq e
} | p_{e,T'}(\ell)| 
&  \leq \frac{1}{2} \sum_{\ell \in \mathrm{leaf}(\mathcal{T})} \sum_{
 T' \neq e
}  \big| W_{e, T'}\big| \; 
        \underbrace{\big| \tr \big[ R(T')^{\dagger} \rho_l) \big] \big|}_{\leq 1} \; \big| \tr \big[ M_l   \big] \big| \\
        \nonumber
        & \leq \frac{1}{2} \sum_{
 T' \neq e
}  \big| W_{e, T'}\big| \; 
         \underbrace{\sum_{\ell \in \mathrm{leaf}(\mathcal{T})} 
        \big| \tr \big[ M_{\ell} \big] \big|}_{=2^{nt}} \\
        \nonumber
        & \leq 2^{O(t^2)}\cdot 2^{-n(t+1)+O(t^{2})} \cdot 2^{nt}\leq2^{-n+O(t^{2})}.
        \nonumber
\end{align}
Here, we have used $\left|W_{e,T'}\right|\leq2^{-n(t+1)+O(t^{2})}$. 
Furthermore, the bound $\big| \tr \big[ R(T')^{\dagger} \rho_l) \big] \big|\leq1$ for all $T'\neq e$ follows from our \cref{thm:unitary-partial-transposes-rT} about unitary partial transposes. In particular, we can apply \cref{cor:tr_rt_rho_ppt_bound}, since
 $\rho_l =\otimes_{i=1}^t\rho^{u_{i-1}}$ is a product state and hence PPT.

\paragraph{Third term: the remaining entries with $T\neq e$.}

The final term proceeds by a similar calculation
that gives
\begin{align}
    \frac{1}{2} \sum_{\ell \in \mathrm{leaf}(\mathcal{T})} \sum_{ T\neq e,T'} | p_{T,T'}(\ell)| 
& \leq \frac{1}{2} \sum_{\ell \in \mathrm{leaf}(\mathcal{T})} \sum_{ T\neq e,T'} \big| W_{T, T'}\big| \; 
        \underbrace{\big| \tr \big[ R(T')^{\dagger} \rho_l) \big] \big|}_{\leq 1} \; \big| \tr \big[R(T) M_l   \big] \big| \\
        \nonumber
        &\leq \frac{1}{2} \sum_{ T\neq e,T'} \big| W_{T, T'}\big|\; \sum_{\ell \in \mathrm{leaf}(\mathcal{T})} 
        \big| \tr\big[R(T) M_l  \big] \big| .
\end{align}
Here, $\big| \tr \big[ R(T')^{\dagger} \rho_l) \big] \big|$ was again bounded via \cref{cor:tr_rt_rho_ppt_bound} as for the second term. To bound $\sum_{\ell \in \mathrm{leaf}(\mathcal{T})} \big| \tr\big[R(T) M_l \big] \big| $ for $T\neq e$, we use the following fact:

\begin{fact}[Duality of trace norm and operator norm for POVMs]\label{fact:duality-trace-operator-norm-povm}
Let $\{M_s\}_s$ be a POVM, i.e., a collection of positive semidefinite operators $M_s \succeq 0$ and $\sum_s M_s = I$. Then, for any operator $A$,
\begin{equation}
    \sum_s \big| \tr \big[A\; M_s   \big] \big| \leq \lVert A\rVert_1. 
\end{equation}
\end{fact}
\begin{proof}
Consider the duality of the trace norm and the operator norm, $\lVert A\rVert_1 = \mathrm{sup}_{\lVert B\rVert_{\infty} \leq 1} |\tr(AB)|$. The claim then follows using that $ \lVert \sum_s \sigma_s M_s \rVert_{\infty}\leq 1$ for any choice of $\sigma_s\in \{1,-1\}$ and writing
 \begin{equation}
      \sum_s \big| \tr \big[A\ M_s   \big]\big| = \underset{\sigma_s\in \{1,-1\}}{\mathrm{sup}} \sum_s \sigma_s \tr \big[A M_s   \big] = \underset{\sigma_s\in \{1,-1\}}{\mathrm{sup}}  \tr \big[A \sum_s\sigma_s M_s   \big] \leq \mathrm{sup}_{\lVert B\rVert_{\infty} } |\tr(AB)|.
 \end{equation}
\end{proof}
        
We can combine \cref{fact:duality-trace-operator-norm-povm} with \cref{thm:non-unitary-partial-transposes}. In particular, by \cref{thm:non-unitary-partial-transposes}, for all $T\neq e$, there exists $S\subseteq [t]$, such that $\lVert R(T)^{\Gamma_S}\rVert_1\leq 2^{n(t-1)}$. 
Hence, using that $\{M_{\ell}\}_{\ell \in \mathrm{leaf}(\mathcal{T})}$ is a POVM and that each $M_{\ell}$ remains a POVM element under partial transposes (because they are product operators), we have for all $T\in \Sigma_{t,t}$ and all $S \subset [t]$
\begin{equation}
    \sum_{\ell \in \mathrm{leaf}(\mathcal{T})} 
        \big| \tr\big[R(T) M_l  \big] \big| =  \sum_{\ell \in \mathrm{leaf}(\mathcal{T})} 
        \big| \tr\big[R(T)^{\Gamma_S} M_l  \big] \big|  \leq\lVert R(T)^{\Gamma_S}\rVert_1,
\end{equation}
using Hölder's inequality. So, in particular, for all $T\neq e$, we have
\begin{equation}
      \sum_{\ell \in \mathrm{leaf}(\mathcal{T})} 
        \big| \tr\big[R(T) M_l  \big] \big| \leq \min_{S\subset\left[t\right]} \lVert R(T)^{\Gamma_S}\rVert_1\leq 2^{n(t-1)} .
\end{equation}
 Overall, the third term is thus bounded as  
\begin{align}
    \frac{1}{2} \sum_{\ell \in \mathrm{leaf}(\mathcal{T})} \sum_{ T\neq e,T'} | p_{T,T'}(\ell)|
    &\leq 
 \frac{1}{2} \sum_{ T\neq e,T'} | W_{T, T'}| \cdot 2^{n(t-1)} \\
 \nonumber
         & \leq2^{O(t^{4})}\cdot  2^{-nt}\left(1+2^{-n+O(t^{2})}\right) \cdot 2^{n(t-1)}=2^{-n+O(t^{4})},
         \nonumber
\end{align}
where we have used $|W_{T,T'}|\leq\left|W_{T,T}\right|\leq2^{-nt}\left(1+2^{-n+O(t^{2})}\right)$.

Since the last term dominates with a scaling of $2^{-n+O(t^{4})}$, we find the TV distance bound claimed in \cref{thm:tv-distance-auxiliary-free}.
\end{proof}

\subsection{The issue with bounding auxiliary-assisted algorithms}\label{ssec:auxiliary-assisted-lower-bound}
Now we turn to auxiliary-assisted strategies, that is, we allow the distinguishing algorithm to operate on $\mathcal{H}_{\mathrm{main}}\otimes \mathcal{H}_{\mathrm{aux}}$. Note that the unknown channel $\mathcal{E}$ only operates on $\mathcal{H}_{\mathrm{main}}$. Without loss of generality, we can assume that the auxiliary system $\mathcal{H}_{\mathrm{aux}}$ is at most the size of $\mathcal{H}_{\mathrm{main}}$, i.e.,  we can take $\mathcal{H}_{\mathrm{aux}}=(\mathbb{C}^{2})^{\otimes n}$ . In analogy to \cref{ssec:auxiliary-free-lower-bound}, we again write out the leaf probabilities under the two hypotheses.
For the random Clifford unitary channel, we find on the one hand
\begin{align}
  \underset{C\sim \Cln}{\mathbb{E}}  p_{\mathcal{C}}(\ell) &=  \underset{C\sim \Cln}{\mathbb{E}}
        \prod_{i=1}^t 
        \tr
        \Big[\, M^{u_{i-1}}_{s_i}\,
        (C \otimes I)\rho^{u_{i-1}}(C^{\dagger} \otimes I) \,\Big]. \\
        \nonumber
        & = \sum_{T,T'\in \Sigma_{t,t}} W_{T, T'}\;  
        \tr_{\mathrm{aux}}
        \Big[ \tr_{\mathrm{main}} \big[ R(T')^{\dagger} \rho_{\ell} \big]\; \tr_{\mathrm{main}} \big[R(T) M_{\ell}  \big] \Big] \\
        \nonumber
         &= \sum_{T,T'\in \Sigma_{t,t}} W_{T, T'}\; \vcenter{\hbox{\includegraphics[scale=1.0]{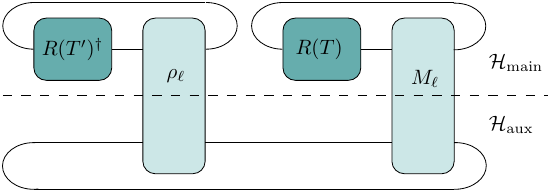}}} \\
         \nonumber
        &:= \sum_{T,T'\in \Sigma_{t,t}}  p_{T,T'}(\ell).
        \nonumber
\end{align}
On the other hand, for the completely depolarizing channel, we find
\begin{align}
     p_{\mathcal{D}}(\ell) &= \frac{1}{2^{nt}}\prod_{i=1}^t 
        \tr
        \Big[\, M^{u_{i-1}}_{s_i}\, \big(I \otimes \tr_{\mathrm{main}} (\rho^{u_{i-1}})\big)
         \,\Big].\\
         \nonumber
        &= \frac{1}{2^{nt}}\tr_{\mathrm{aux}}
        \Big[ \tr_{\mathrm{main}} \big[ \rho_{\ell} \big]\; \tr_{\mathrm{main}} \big[M_{\ell} \big] \Big] \\
        \nonumber
        & =  \frac{1}{2^{nt}} \vcenter{\hbox{\includegraphics[scale=1.0]{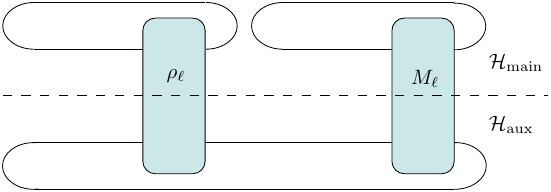}}}
\end{align}

Based on these expressions and their corresponding diagrammatic versions, we now explain why our current proof strategy from the auxiliary-free case (see \cref{ssec:auxiliary-free-lower-bound}) does not seem to generalize to the auxiliary-assisted setting.

For the auxiliary-free bound, our strategy was to bound bound separately the contributions $ \sum_{\ell \in \mathrm{leaf}(\mathcal{T})}  |p_{T,T'}(\ell)| $ for each pair $T,T' \in \Sigma_{t,t}$ (see also \cref{eq:tv-distance-expansion}).
The main idea for doing this was transforming these contributions by taking partial transposes of the $R(T), R(T')$, the generators of the Clifford commutant. Crucially, in the auxiliary-free setting, we were allowed to choose partial transposes independently for $R(T)$ and $R(T')$. In particular, we could choose two independent subsets $S, S' \subset[t]$ to transform $R(T)$ to $R(T)^{\Gamma_{S}}$ and $R(T')$ to $R(T')^{\Gamma_{S'}}$, to get
\begin{align}
   \tr \big[ R(T') \rho_{\ell} \big]\; \tr \big[R(T) M_{\ell}  \big] 
   &= \tr \big[ R(T')^{\Gamma_{S'}} \rho_{\ell}^{\Gamma_{S'}} \big]\; \tr \big[R(T)^{\Gamma_{S}}  M_{\ell}^{\Gamma_{S}}  \big], \\
   \vcenter{\hbox{\includegraphics[scale=1.0]{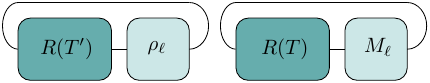}}} 
   &= \vcenter{\hbox{\includegraphics[scale=1.0]{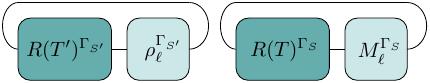}}}. \nonumber
\end{align}
Importantly, since $\rho_{\ell}$ and $M_{\ell}$ are product operators, they have the PPT property, and thus both $\rho_{\ell}^{\Gamma_{S'}}$ and $M_{\ell}^{\Gamma_{S}}$ remain positive semi-definite for all choices of $S, S' \subset [t]$.
Then, both trace terms could be bounded independently via norm inequalities such as Hölder's inequality.

On the other hand, in the auxiliary-assisted setting, the contribution for the pair $T,T'$ involves another additional trace over $\mathcal{H}_{\mathrm{aux}}$,
\begin{align}
        \tr_{\mathrm{aux}}
        \Big[\! \tr_{\mathrm{main}} \big[ R(T')^{\dagger} \rho_{\ell} \big]\ \tr_{\mathrm{main}} \big[R(T) M_{\ell}  \big] \Big] 
         = \vcenter{\hbox{\includegraphics[scale=1.0]{images/p_T_T.pdf}}}.
\end{align}

Also, $\rho_{\ell}$ and $M_{\ell}$ are no longer positive semi-definite under arbitrary partial transposes over the $2t$ copies of $\big(\mathcal{H}_{\mathrm{main}}\otimes \mathcal{H}_{\mathrm{aux}}\big)^{\otimes t} = \big(\mathbb{C}^{2^n}\big)^{\otimes 2t}$. Instead, they only have the PPT property under partial transposes that act on the same copies in both the main and auxiliary 
spaces. Concretely, this means
\begin{equation}
\rho_{\ell}^{\Gamma_S}, \; M_{\ell}^{\Gamma_S} \geq 0 
\quad \text{for all } S \subset [2t] \text{ of the form } 
S = S' \cup (S'+t), \; S' \subset [t].
\end{equation}
This reflects that in each round of the adaptive algorithm, the input state and measurement may act jointly on $\mathcal{H}_{\mathrm{main}}\otimes \mathcal{H}_{\mathrm{aux}}$.
Combining these two considerations, we find that it is no longer possible to independently choose partial transposes of $R(T)$ and $R(T')$ without incurring extra dimensional factors. However, this has crucially been necessary in bounding these contributions in the previous section.

\appendix
\section{Algorithmic proof of \texorpdfstring{\Cref{thm:unitary-partial-transposes-rD}}{Theorem~\ref{thm:unitary-partial-transposes-rD}}}\label{sec:algorithmic-proof}
In this appendix we provide a proof of \Cref{thm:unitary-partial-transposes-rD} that is explicitly algorithmic and uses only elementary linear algebra (which makes it somewhat more unwieldy than the matroid-theoretic proof in the main text). 

\unitarypartialtransposes*

\begin{proof}
Consider a generator matrix $M_D = [A_D|B_D]$ of the space $D$. We can obtain a generator matrix $M_{D^S}$ of $D^S$ by swapping the columns $[A_D]_i$ with $i \in S$ with the corresponding columns $[B_D]_i$. Note that $|D_{LD}^S|  = |\mathrm{ker}\p{A_{D^S}}|$. The goal is thus to find, given $M_{D}$, a set of transpositions $S$ such that $A_{D^S}$ is invertible. Since partial transpositions compose, we can do this sequentially. We will drop the subscripts from the matrices $A,B,M$ when they are clear from context.
We will use $A_{*,i}$ to denote the $i$-th column of $A$ and $B_{j,*}$ to denote the
  $j$-th row of $B$.
In addition, we will use $A_{\le k, \le k}$ to denote the up-left $k\times k$
  submatrix of $A$, and define $A_{>k,>k}$, $A_{\le k, >k}$ and $A_{>k, \le k}$
  similarly.

Because elementary row operations commute with partial transpositions and they do not change the 
  rank of $A$ or $B$, we can apply them freely to the generator matrix $M$.
In the following algorithm, we transform $A$ into a full rank matrix by swapping the 
  corresponding columns and applying elementary row operations to $M$.

\begin{algorithm}[ht]
\caption{\textsc{Unitary-Partial-Transpose}($t, M$)}
\SetKwInOut{Input}{Input}
\SetKwInOut{Goal}{Goal}
\Input{A natural number $t$ and a generator matrix $M=[A|B] \in \mathbb{F}_2^{t\times 2t}$}
\Goal{Transform $A$ into the identity matrix $I_{t\times t}$ by swapping the
  corresponding columns and applying elementary row operations}
\label{alg:unitary-partial-transpose}
\For{$k = 0, 1, \dots, t-1$}{
  Let $t_0 \gets \textsc{Reduce-Augmenting-Path}(k,t,M)$\; 
      \tcc{Now we have $k < t_0 \le t,\; B_{t_0,k+1}=1$}
  Swap columns $A_{*,k+1}$, $B_{*,k+1}$, and swap rows $M_{t_0,*}$, $M_{k+1,*}$ 
  to make $A_{k+1,k+1}=1$\;
  Eliminate all other $1$’s in $A_{*,k+1}$ by adding row $A_{k+1,*}$ to the other rows\;
}
\end{algorithm}

Here the subroutine \textsc{Reduce-Augmenting-Path}($k,t,M$) always outputs $k<t_0\le t$ and
  guarantee $B_{t_0,k+1}=1$. 
In order to find such $t_0$ it may also applies partial transpositions and elementary row operations
  to $M$.
We will introduce \textsc{Reduce-Augmenting-Path}($k,t,M$) and prove its correctness by induction.

\paragraph{Base case ($k=0$):} Note that the first column $B_{*,1}$ of $B$ always contains at least one non-zero entry. Otherwise $M$ would generate a subspace of dimension $t$ of the space $\mathbb{F}_2^{2t-1}$. Since $t>(2t-1)/2$, this space can not be self-orthogonal, which is a contradiction. 

\paragraph{Induction step ($1\le k < t$):} Now assume $M$ is of the form
\begin{equation}
M = \left[\begin{array}{cc|cc} I_{k\times k} & A_{\leq k, >k} &B_{\leq k, \leq k} & B_{\leq k, >k}\\
0 & A_{>k, >k} & B_{>k,\leq k} & B_{>k,>k}\end{array}\right],
\end{equation}
 for some $1\leq k< t$.

We define a simple directed graph $G=(V=[k+1],E)$ such that 
  for any $i,j\in [k+1]$, $(i,j)\in E$ if and only if
  $B_{j, i}=1$.
Let $V_0\subseteq V$ be the vertex set $\{i\in V\mid B_{>k,i}\neq 0\}$.
We will first argue that there must exist a path from $k+1$ 
  to some vertex in $V_0$.
If $k+1\in V_0$, we can directly complete the induction step,
  by moving the $1$ to $A_{k+1,k+1}$ and eliminating all
  other $1$'s in the column $A_{*,k+1}$.
Otherwise, we can take a shortest path from $k+1$ to $V_0$ 
  and iteratively make changes to $M$ to reduce the 
  length of this path by $1$ until $k+1\in V_0$.

\paragraph{There exists a path from $k+1$ to $V_0$:}
Assume by contradiction that there is no path in $G$
  from $k+1$ to $V_0$.
Let $L \subseteq V$ be the set of all vertices reachable 
  from $k+1$, including $k+1$ itself.
Since there is no directed edge from any vertex in $L$ to 
  $V\setminus L$, we have that $B_{j', j}=0$ for all 
  $j\in L, j'\in V\setminus L$.
In addition, since $L\cap V_0 = \varnothing$ by assumption,
  for 
  any $j\in L$, we have $B_{>k, j}=0$.

We now focus on the $2t-k-|L|$ columns $A_{*,>k}$, 
  and $B_{*, [t]\setminus L}$.
Let $C$ be the set of indices of these columns in $M$.
Now consider a subset of the rows of $M_{*,C}$, in particular 
  we consider only the rows in the set 
  $[t]\setminus (L\setminus \{k+1\})$, which is of 
  size $t-|L|+1$.
Let $w_1,w_2,\dots,w_{k-|L|+1}\in \mathbb{F}_2^{|C|}$ be
  these row vectors and let $W = \operatorname{Span}\left(w_1,\dots,w_{k-|L|+1}\right)$.
Let $v_1,v_2,\dots,v_{t-k}\in \mathbb{F}_2^{|C|}$ be
  the row vectors of $M_{>k, C}$, and 
  let $V = \operatorname{Span}\left(v_1,\dots,v_{t-k}\right)$.

We will first show that $w_1, w_2, \dots, w_{k-|L|+1}$ 
  are linearly independent and that 
  $v_1, v_2, \dots, v_{t-k}$ are also linearly 
  independent. 
Therefore, $\dim W = k-|L|+1$ and $\dim V = t-k$.
Then we will argue that $V$ is self-orthogonal, meaning 
  $V\subseteq V^{\perp}$ and $V$ is orthogonal
  to $W$, meaning $V\subseteq W^{\perp}$.
We will further show that $V\cap W = \{0\}$.
Combining all these facts, we can prove using 
  dimension inequalities that $\dim V < t-k$, 
  which contradicts $\dim V = t-k$.

We will use the fact that the row vectors of $M$ are 
  linearly independent and $D^S$ is self-orthogonal.
Note that $A_{\le k, \le k} = I_{k,k}$ and 
  $B_{[k]\setminus (L\setminus \{k+1\}), L} = 0$. 
For the vectors $w_1, w_2, \dots, w_{k-|L|+1}$, 
  we have that for all $1 \le i \le k-|L|+1$, 
  $w_i^T w_i = 1$, and for all 
  $1 \le i \neq j \le k-|L|+1$, $w_i^T w_j = 0$.
To see that $w_1,w_2,\dots,w_{k-|L|+1}$ are linearly
  independent, assume there exist coefficients
  $\{\alpha_i\}_{i=1}^{k-|L|+1}$ such that
  $\sum_{i=1}^{k-|L|+1}\alpha_iw_i=0$.
Then for all $1\le i\le k-|L|+1$, we have that $\alpha_i=
  \left(\sum_{j=1}^{k-|L|+1}\alpha_jw_j\right)^Tw_i=0$.
This means that $w_1,w_2,\dots,w_{k-|L|+1}$ are linearly
  independent.
For $v_1,v_2,\dots,v_{t-k}$, we note that $A_{>k, \le k}=0$
  and $B_{>k, L} = 0$; therefore $v_1,v_2,\dots,v_{t-k}$ 
  must be linearly independent since $M$ is full rank.
We conclude that $\dim W =k-|L|+1 $ and $\dim V = t-k$.

Since $D^S$ is self-orthogonal, it is easy to see that
  $V\subseteq V^{\perp}$ and $V\subseteq W^{\perp}$.
To show $V\cap W = \{0\}$, we will show that
  $W\cap W^{\perp} = \{0\}$.
Assume there exist coefficients
  $\{\alpha_i\}_{i=1}^{k-|L|+1}$ such that
  $\sum_{i=1}^{k-|L|+1}\alpha_iw_i\in W^{\perp}$.
Then for all $1\le i\le k-|L|+1$, we have that $\alpha_i=
  \left(\sum_{j=1}^{k-|L|+1}\alpha_jw_j\right)^Tw_i=0$.
Therefore, $\sum_{i=1}^{k-|L|+1}\alpha_iw_i=0$.
This means that $W\cap W^{\perp} = \{0\}$.
We conclude that $V\cap W = \{0\}$.

Now we are ready to obtain a contradiction regarding
  the dimension of $V$.
Note that $\dim (V+W)+\dim (V+W)^\perp = 2t-k-|L|$.
Since $V\cap W=\{0\}$, we have that
  $\dim (V+W) = \dim V + \dim W$.
Since $V\subseteq V^{\perp}$ and $V\subseteq W^{\perp}$,
  we have that $V\subseteq (V+W)^{\perp}$, and therefore
  $\dim V\le \dim (V+W)^{\perp}$.
To conclude, we have that $\dim V+\dim W+\dim V\le 2t-k-|L|$,
  which simplifies to $2\dim V\le 2t-2k-1$.
This implies $\dim V < t-k$.
A contradiction follows because we already know that
  $\dim V=t-k$.
To summarize, there must exist a path from $k+1$ to $V_0$.

\paragraph{Length zero path, which is $k+1 \in V_0$:}
If $k+1\in V_0$, 
  we know that there exists a $t_0>k$ such that $B_{t_0,k+1}=1$.
We can then complete the induction step by swapping the 
  columns $A_{*,k+1}$ and $B_{*,k+1}$, which moves the $1$ 
  to $A_{k+1,k+1}$. We then eliminate all other $1$'s in 
  the column $A_{*,k+1}$ by row operations with $A_{k+1,k+1}=1$.
\begin{algorithm}[t]
\caption{\textsc{Reduce-Augmenting-Path}($k,t, M$)}
\SetKwInOut{Input}{Input}
\SetKwInOut{Goal}{Goal}
\Input{Natural numbers $k$, $t$ and a generator matrix $M=[A|B] \in \mathbb{F}_2^{t\times 2t}$}
\Goal{Find $k<t_0\le t$ and guarantee $B_{t_0,k+1}=1$, by swapping corresponding columns
  and applying elementary row operations}
\label{alg:reduce-augmenting-path}
Construct directed graph $G=(V=[k+1],E)$ and $V_0\subseteq V$ as described above\;
Find the shortest path with length $l$, $a_0=k+1, a_1, \dots, a_l \in V_0$ from $k+1$ to $V_0$\;
There exists $k < t_0 \le t$ such that $B_{t_0,a_l}=1$\;

\For{$i = l, l-1, \dots , 1$}{
   Eliminate all other $1$’s in the column $B_{*,a_i}$ using $B_{t_0,a_i}=1$ via row operations\;
   Swap columns $A_{*,a_i}$ and $B_{*,a_i}$ via a partial transpose\;
   Swap rows $M_{t_0,*}$ and $M_{a_i,*}$\;
}
\end{algorithm}

\paragraph{Reduce the length of one shortest path:}
Let $a_0=k+1,a_1,a_2,\dots,a_{l-1},a_l\in V_0$ be a shortest path from $k+1$ to $V_0$. 

If $l=0$, then $k+1\in V_0$ and we can complete the induction step by the argument above.
On the other hand, if $l>0$, we will perform the following three-step subroutine to reduce the path length $l$ by exactly $1$. Note that this changes the graph $G$.

\begin{enumerate}
    \item Eliminate all other $1$'s in the column
      $B_{*,a_l}$ using $B_{t_0,a_l}=1$ via row operations.
    \item Swap columns $A_{*,a_l}$ and $B_{*,a_l}$,
      via a partial transpose.
    \item Swap rows $M_{t_0,*}$ and $M_{a_l,*}$.
\end{enumerate}

After the first step, we have that
  $B_{*,a_l}=0$ except $B_{t_0,a_l}=1$.
Note that this step does not change $A_{*, \le k}$
  or $B_{*,a_i}$ for any $0\le i\le l-1$, because
  $A_{t_0, \le k}=0$ and $B_{t_0,a_i}=0$ 
  for all $0\le i\le l-1$, since $a_i\notin V_0$ for 
  all $0\le i\le l-1$.
After the second step, we have that $A_{t_0,a_l}=1$.
By the definition of the graph $G$, because 
  there is an edge from $a_{l-1}$ to $a_l$, 
  we also have $B_{a_l,a_{l-1}}=1$.
After the third step, we have that $A_{a_l,a_l}=1$
  and $B_{t_0, a_{l-1}}=1$.
  
We will argue that after these three steps, 
  the sequence $a_0 = k+1, a_1, \dots, a_{l-1}$
  is a path from $k+1$ to $V_0$.
It is easy to see that $a_{l-1} \in V_0$ since
  $B_{t_0, a_{l-1}}=1$.
It is sufficient to show that for all $0\le i\le l-2$,
  $B_{>k,a_i}=0$.
Since the path $a_0,a_1,\dots,a_l$ is a shortest path, 
  for all $0\le i\le l-2$, we have that $B_{a_l,a_i}=0$.
Therefore, after the third step (swapping the rows),
  for all $0\le i\le l-2$, $B_{>k,a_i}$ remains $0$.
Hence $a_0,a_1,\dots,a_{l-1}$ is a shortest path
  of length $l-1$ from $k+1$ to $V_0$.
  
\end{proof}

\bibliographystyle{alphaurl}
\bibliography{refs}

\end{document}